\documentclass[draft,onecolumn]{IEEEtran}
\usepackage{Template}

\usepackage[utf8]{inputenc}

\title{Typical Error Exponents: \\A Dual Domain Derivation}
\author{Giuseppe Cocco, Albert Guill\'en i F\`{a}bregas and Josep Font-Segura
		\thanks{G. Cocco and J. Font-Segura are with Department of Information and 
			Communication
			Technologies, Universitat Pompeu Fabra, Barcelona 08018, Spain (e-mail: giuseppe.cocco@upf.edu, josep.font@upf.edu).
			A.~Guill\'en i F\`abregas is with the  Department of Engineering, 
			University
			of Cambridge, Cambridge CB2 1PZ, U.K. and the Department of Information and 
			Communication
			Technologies, Universitat Pompeu Fabra, Barcelona 08018, Spain (e-mail: guillen@ieee.org). 
			
			This work has been funded in part by the European Research Council under ERC grant agreement 725411, by the Catalan Secretary of Universities and Research under a Beatriu de Pin\'{o}s postdoctoral fellowship, and by the European Union's Horizon 2020 research and innovation programme under the Marie Sk{\l{}}odowska-Curie grant agreement 801370.
			
Part of this work has been presented at the IEEE International Symposium on Information Theory (ISIT) 2021, Melbourne, Australia.
		}
	}

\begin{document}
\maketitle

\begin{abstract}
This paper shows that the probability that the error exponent of a given code randomly generated from a pairwise-independent ensemble being smaller than a lower bound on the typical random-coding exponent tends to zero as the codeword length tends to infinity. This lower bound is known to be tight for i.i.d. ensembles over the binary symmetric channel and for constant-composition codes over memoryless channels. Our results recover both as special cases and remain valid for arbitrary alphabets, arbitrary channels---for example finite-state channels with memory---, and arbitrary pairwise-independent ensembles. We specialize our results to the i.i.d., constant-composition and cost-constrained ensembles over discrete memoryless channels and to ensembles over finite-state channels.
\end{abstract}

%%%%%%%%%%%%%%%%%%%%%%%%%%%%%%%%%%%
\section{Introduction}
Since Shannon's work \cite{shannon1948mathematical}, the random coding method remains a key tool in information theory and is applied to a wide variety of communication and compression settings. Specifically, by analysing the probability of error averaged over a certain ensemble of randomly generated codes, it is possible to show the existence of sequences of codes with a certain rate having vanishingly small probability of error. 

A refinement of the random coding analysis allows to show that, for any \ac{DMC}, the ensemble average error probability decays exponentially according to a certain error exponent, often referred to as the random-coding exponent \cite{fano1961transmission,gallagerBook}. The resulting exponent is known to be tight at high rates \cite{sgb}, but not at low rates. By expurgating poor codebooks from the ensemble, it is possible to show that there exist sequences of codes that attain the expurgated exponent, known to be strictly better than the random coding one at low rates \cite{gallagerBook, csiszar1977new, czisar_TIT1981}. While it is difficult to infer the structure of the resulting expurgated ensemble, it is useful to prove the existence of good low-rate codes. The expurgated exponent is known to be tight at rate that tends to zero \cite{sgb}. It is also possible to show the existence of codes that attain the maximum of the expurgated and random-coding exponents \cite{czisar_TIT1981} and to construct explicit random-coding ensembles that attain the same maximum \cite{generalizedGV_fabregas2020}. 

As opposed to the exponent of the ensemble-average error probability, Barg and Forney \cite{barg_forney_TIT2002} studied the average error exponent attained by the i.i.d. random-coding ensemble over the binary symmetric channel (BSC). They coined the resulting exponent the typical error exponent and showed that at low rates, it can be strictly higher than the random-coding exponent and strictly lower than the expurgated ---the typical and expurgated error exponents coincide for rate zero. Nazari {\em et al.} \cite{errExp_MAC_LB_nazari_TIT2014} derived bounds to the typical error exponent with constant-composition ensembles for discrete-memoryless and multiple-access channels. Significant progress in the understanding of typical error exponents has been made since. Merhav has derived the typical error exponent in a number of settings, including DMCs with constant-composition codes \cite{merhav_TIT2018}, trellis codes \cite{merhav2019error}, and power-constrained ensembles over the coloured Gaussian noise channel \cite{merhav2019errorc}. Recently, Tamir {\em et al.} \cite{LargeDevLogErrProb_tamir_TIT2020} showed that the probability of finding a code whose exponent is lower (higher) than the typical error exponent decays exponentially (resp. double-exponentially) illustrating an asymmetry in the lower and upper tails of the distribution of the error exponent of random constant-composition codes. The concentration properties of the error exponent of randomly generated codes is studied in \cite{Truong}. Tamir and Merhav \cite{tamir2022universal} have shown that the typical error exponent can be achieved by a universal decoder ignorant of the channel law. Specifically, they have shown that a stochastic decoder based on the empirical mutual information between the received sequence and each codeword attains the typical error exponent. Most aforementioned derivations of the typical error exponent are done in the primal domain, i.e., they minimize a certain objective function involving information quantities such as the mutual information, entropy or the relative entropy over a constrained set of probability distributions.

In this paper, we refine a Lemma by Gallager to provide a derivation of a lower bound to the typical error exponent for general channels and pairwise-independent random-coding ensembles. As opposed to most previous works, our derivation is naturally a dual-domain derivation \cite{gallagerBook}, i.e., it is expressed as a maximization of an exponent function over certain non-negative real-valued parameters. This usually has the advantage that any choice of parameters provides an achievable exponent, and that it naturally extends to arbitrary alphabets. In \cite{merhav_lagrange_dual_TRC_2020} Lagrange duality was applied to the primal domain expression to obtain a dual domain expression for a lower bound on the typical error exponent for \ac{i.i.d.} random coding and mismatched stochastic decoding. The overall expression involves an optimization over five parameters. Unlike \cite{merhav_lagrange_dual_TRC_2020}, we directly derive a dual domain achievability proof of the typical error exponent.
Our lower bound is valid for any channel and code ensemble with pairwise-independent codewords, thus significantly extending previous results, including \cite{merhav_lagrange_dual_TRC_2020}. We particularize our result to i.i.d., constant composition and cost-constrained ensembles over DMCs and we show that these coincide with well known expressions for the typical error exponent for DMCs, i.e., is tight in such cases. As a consequence of our derivation, we provide a rather direct dual-domain derivation of the expurgated exponent for constant composition ensembles.
We also particularize our bound to the case of \ac{FSC}. Although it is known that reliable communication over a finite-state channel is theoretically possible at any rate below capacity \cite[pp.~176--182]{gallagerBook}, comparatively little research has been carried out with respect to \ac{DMC} \cite{MushkinExp_GilbEll_TIT89}. 
As pointed out in \cite{zero_rate_memory_MerhavTIT2015}, the most studied channel model with memory has been the finite-state channel (FSC) and some of its special cases. The channel coding theorem for finite-state indecomposable channels was first proved by Blackwell, Breiman and Thomasian \cite{blackwell_exp_FSC_math_amm_58}. The random-coding exponent for \ac{FSC} was studied in  \cite{blackwell_Exp_FSC_1961}  and \cite{Yudkin_ISIT67} and further developed in \cite[Sec.~5.9]{gallagerBook}. 
In this paper, we  provide an expression for the corresponding lower bound on the typical error exponent. Finally, we present an extension of our results to the case of a mismatched decoder, of which the ML is a special case, which is valid for all channels and pairwise-independent code ensembles.

The paper is structured as follows. Section \ref{sec:prelim} introduces the preliminaries and the main notation used in the paper. Section \ref{sec:main_res} introduces the refinement of Gallager's lemma and derives the main results of the paper. Section \ref{sec:spec_ensembles} particularizes the results of Section \ref{sec:main_res} to memoryless channels with i.i.d., constant-composition and cost-constrained code ensembles. Section \ref{sec:memory} shows how to adapt the main results in \ref{sec:main_res} to finite-state channels. Proofs of some lemmas can be found in the Appendix.

%%%%%%%%%%%%%%%%%%%%%%%%%
\section{Preliminaries}
\label{sec:prelim}
We consider coding over discrete channels with conditional probability distribution $\Wnvec$, being $\x\in\Xc^n$ and $\y\in\Yc^n$ the  transmitted and received sequences of length $n$, and $\Xc,\Yc$ the finite channel input and output alphabets, respectively. For memoryless channels we have $\Wnvec=\prod_{i=1}^nW(y_i|x_i)$, where $x_i\in\Xc, y_i\in\Yc$. A code $\Cns=\{\x_1,\ldots,\x_{M_n}\}$ is a set of $M_n$ codewords of length $n$. We consider a maximum-likelihood decoder that outputs message estimate $\hat m$ as
\beq
\hat m = \argmax_{m\in\{1\dotsc M_n\}} W^n(\y|\x_m).
\eeq
Assuming equiprobable messages, the error probability of a fixed code $\Cns$ is given by 
\beq
\Pecns  =  \frac{1}{M_n}\sum_{m=1}^{M_n}\Pecnm,
\eeq
where $\Pecnm$ is the error probability conditioned to codeword $\x_m$ being transmitted. We define the finite-length error exponent of a code $\Cns$ as 
\begin{equation}
	\Ens  =  -\frac 1n \log \Pecns.
	\label{eq:error1}
\end{equation}
An exponent $E$ is said to be achievable when there exists a sequence of codes $\{\Cns\}_{n=1}^{\infty}$ such that $\liminf_{n\to\infty} \Ens \geq E$.

Lower bounds on the error exponent of codes used over discrete memoryless channels are traditionally derived using random-coding arguments \cite[Sec.~5.6]{gallagerBook}, \cite[Ch.~10]{csiszar2011information}. More specifically, let $\Cn$ be the random variable representing a code randomly generated according to some probability distribution. In code ensembles with pairwise-independent codewords, such as the i.i.d., the constant-composition or the cost-constrained ensembles later discussed in Sec.~\ref{sec:spec_ensembles}, the $M_n$ codewords are generated independently with some probability distribution $Q^n(\x)$. For such ensembles and sufficiently large $n$, there exists a sequence of codes $\{\Cns\}_{n=1}^{\infty}$ whose limiting error exponent, given in~\eqref{eq:error1} as $n\to\infty$, is at least as large as the random-coding error exponent $\Er$ given by
\beq
\Er =  \lim_{n\to\infty} -\frac{1}{n}\log  \EE[\Pecn],
\label{eq:rc}
\eeq
where $
R  =  \lim_{n\to\infty} \frac 1n \log M_n$ is the code rate and $Q$ is the asymptotic single-letter version of $Q^n$. For the i.i.d., constant composition and cost-constrained ensembles over DMCs, \eqref{eq:rc} is the actual error exponent of the ensemble average error probability, and not only a lower bound.
%, Sec.~\ref{sec:spec_ensembles}
The random-coding error exponent in~\eqref{eq:rc} is known to be tight at high rates \cite{sgb} while it is not at low rates. Expurgation allows to show the existence of a codebook with an improved exponent, the expurgated exponent $\Eex$ \cite[Eq.~(5.7.11)]{gallagerBook}.

In contrast to the limiting exponent of the ensemble-average error probability, Barg and Forney in~\cite{barg_forney_TIT2002} defined the typical random-coding error exponent $\Etrc$ as the limiting expected error exponent of the ensemble, that is
\beq
\Etrc  =  \lim_{n\to\infty} -\frac 1n \EE[\log P_{\rm e}(\Cn)].
\label{eq:trc1}
\eeq
The typical error exponent improves over the random-coding error exponent~\eqref{eq:rc} at low rates, and is achieved by codes in the specified ensemble, unlike the expurgated exponent, achieved by codes with unknown structure. Therefore, the typical error exponent emerges as the error exponent attained by a given ensemble for a coding length that tends to infinity. Interestingly, the expressions of the typical and expurgated exponents are strongly connected. For the i.i.d.~\cite{barg_forney_TIT2002} and the constant-composition~\cite{merhav_TIT2018},~\cite[Lemma 3]{errExp_MAC_LB_nazari_TIT2014} ensembles, the typical error exponent is shown to satisfy $\Etrc= \Eexx + R\leq \Eex$, with equality for $R=0$. In \cite{merhav_TIT2018} an inequality sign is used, but this is only because the improved expurgated presented in \cite[Sec.~1, point 4]{merhav_TIT2017} is used instead of Gallager's.

%%%%%%%%%%%%%%%%%%%%%%%%%%%%%%%%%%
\section{Main Result}\label{sec:main_res}
The main result of this paper is a consequence the following result, a refinement of a Lemma by Gallager in~\cite[p.~151]{gallagerBook}.

\begin{lemma}
\label{lem:main}
Let $\gamn$ be a sequence in $n$ taking values in $\mathbb{R}^+$. For an arbitrary random-coding ensemble and $s>0$, it holds that
\beq
\PP\Bigl [ \Pecn \geq \gamn^{\frac{1}{s}}\EE[\Pecn^s]^{\frac{1}{s}} \Bigr ]\leq \frac{1}{\gamn}.
\label{eq:4}
\eeq
\end{lemma}

\begin{IEEEproof}
For any $s>0$, consider the random variable $Z=\Pecn^s$ and let $a=\gamma_n\mathbb{E}[Z]$ where $\gamn$ is a positive real-valued sequence. Markov's inequality implies that $\mathbb{P}[Z \geq a] = \mathbb{P}[Z^\frac{1}{s} \geq a^\frac{1}{s}] \leq \frac{\mathbb{E}[Z]}{a}$, obtaining~\eqref{eq:4}.
\end{IEEEproof}

While Lemma~\ref{lem:main} holds for any positive-valued sequence $\gamn$, by adding the constraint that $\gamn$ be monotonically increasing such that $\lim_{n\rightarrow\infty}\gamn=\infty$, we obtain that the probability to randomly generate a code $\Cns$ such that
\beq
\Pecns < \gamn^{\frac{1}{s}}\EE[\Pecn^s]^{\frac{1}{s}}
\label{eq:5}
\eeq
is larger than $1-\frac{1}{\gamn}$, a quantity that tends to $1$ as $n\to\infty$. The r.h.s.~of~\eqref{eq:5} shows a strong connection with the typical random-coding exponent~\eqref{eq:trc1} as $
\lim_{s\to 0} \EE[\Pecn^s]^{\frac{1}{s}}  = \mathbb{E}[\log \Pecn]$. However, since we assume that $\gamma_n>1$ from a certain $n$, the bound~\eqref{eq:5} is tightened for $s<\infty$. This result is subsumed in Theorems~\ref{theo5} and \ref{theo:2}, valid for channels with arbitrary alphabets or memory and pairwise-independent code ensembles. Before moving to Theorem~\ref{theo5} we define the following quantities.

\begin{definition}
We define
\begin{align}\label{def:1}
\Etrclbl  =  \EexnRR + R - \delta_n,
\end{align}
where 
%$\delta_n>0$ such that $\lim_{n\to\infty}\delta_n = 0$,
\begin{align}\label{eqn:expu}
	\Eexn = E_{\rm x}^n(\hat\lambda_n,Q^n) - \hat\lambda_n R
	%\label{eq:18}
\end{align}
is the multi-letter version of the expurgated exponent,
\begin{align}\label{eqn:ex}
\Exnvarrho& =  -\frac{1}{n}\log \biggl( \sum_{\x}\sum_{\x'} Q^n(\x)Q^n(\x') Z_n(\x,\x')^\frac{1}{\lambda}\biggr)^\lambda,
\end{align}
$Z_n(\x,\x')=\sum_{\y}\sqrt{\Wnvec\Wnvecp}$ is the Bhattacharyya coefficient between $\x,\x'\in\Xc^n$,
\begin{align}
\hat\lambda_n = \argmax_{\lambda\geq 1} \bigl\{ E_{\rm x}^n(\lambda,Q^n) - \lambda 2R\bigr\}
\label{eq:rhohatn}
\end{align}
 is the bound parameter that yields the highest exponent, while
 \begin{align}
	\delta_n = \frac{\hat\lambda_n}{n} \log \gamma_n,
	\label{eq:delta}
\end{align}
and $\gamma_n$ is a positive-defined sequence.% such that $\gamma_n \to \infty$.

\end{definition}

\begin{definition}\label{def:2}
Let
\begin{align}
\Etrclbh & =   \Ern - \iota_n,
\end{align}
where
\begin{align}\label{eqn:rce}
\Ern=  E_0(\hat{\rho}_n,Q^n)-\hat{\rho}_n R
\end{align}
is the multi-letter  version of Gallager's random-coding exponent in~\cite[Eq.~(5.6.16)]{gallagerBook},
\begin{align}\label{eq:E0_gen}
E_0^n(\rho,Q^n) =  -\frac{1}{n}\log\bigg(\sum_{\y}\bigg(\sum_{\x}Q^n(\x)\Wnvec^{\frac{1}{1+\rho}}\bigg)^{1+\rho}\bigg),
\end{align}
$\hat\rho_n$ is the optimal bound parameter
\begin{align}
\hat\rho_n = \argmax_{0\leq\rho\leq 1} \bigl\{ E_0(\rho,Q^n)-\rho R\bigr\},
\label{eq:rhohatn_E0}
\end{align}
and
  \begin{align}
	\iota_n = \frac{1}{n} \log \gamma_n,
	\label{eq:delta}
\end{align}
where $\gamma_n$ is a positive-defined sequence.% such that $\gamma_n \to \infty$.

\end{definition}

\begin{theorem}\label{theo5}
For a channel $W^n$ and a pairwise-independent ensemble with codeword distribution $Q^n$, it holds that
\begin{equation}\label{eqn:theo5_statement}
\mathbb{P}\left[ \En > \Etrclbmax \right]\geq 1-\frac{1}{\gamn},
\end{equation}
where $\gamma_n$ is a positive real-valued sequence and 
\begin{align}\label{eqn:trc_lb_def}
\Etrclbmax =  \max\left\{ \Etrclbl,\Etrclbh\right\}.
\end{align}
\end{theorem}

\begin{IEEEproof}
We start deriving an upper bound on the average tilted error probability. 

For a given code $\Cns$ with equiprobable messages, we have: 
\begin{align}\label{eqn:psum_bound_rce}
\Pecns & =  \frac{1}{M_n}\sum_{m=1}^{M_n}\Pecnm\\
&\leq \frac{1}{M_n}\sum_{m=1}^{M_n}\sum_{\y}\Wnvecm^{\frac{1}{1+\rho}}\bigg(\sum_{m'\neq m}\Wnvecmp^{\frac{1}{1+\rho}}\bigg)^{\rho},\label{eqn:psum_bound_rce_1}
\end{align}
where \eqref{eqn:psum_bound_rce_1}, valid for $\rho\geq 0 $, is obtained from Gallager's bound to a given code \cite{gallager1965simple}. 
 Rising \eqref{eqn:psum_bound_rce_1} to the power of $s$ with range in $(0,1]$, we have that:
\begin{align}\label{eqn:psum_bound_rce_s}
\Pecns^s&\leq \Bigg(\frac{1}{M_n}\sum_{m=1}^{M_n}\sum_{\y}\Wnvecm^{\frac{1}{1+\rho}}\bigg(\sum_{m'\neq m}\Wnvecmp^{\frac{1}{1+\rho}}\bigg)^{\rho}\Bigg)^s \\
&\leq \sum_{m=1}^{M}\Bigg(\frac{1}{M_n}\sum_{\y}\Wnvecm^{\frac{1}{1+\rho}}\bigg(\sum_{m'\neq m}\Wnvecmp^{\frac{1}{1+\rho}}\bigg)^{\rho}\Bigg)^s, \label{eqn:psum_bound_rce_s_2}
\end{align}
where \eqref{eqn:psum_bound_rce_s_2} follows from the fact that $(\sum a_i)^s\leq \sum (a_i)^s$ for $s\in(0,1]$ \cite[Ch.~5]{gallagerBook}.
Taking the ensemble average of \eqref{eqn:psum_bound_rce_s_2} we obtain:
\begin{align}\label{eqn:psum_bound_rce_s_mean}
\mathbb{E}[\Pecn^s] &\leq  \EE\left[\sum_{m=1}^{M_n}\Bigg(\frac{1}{M_n}\sum_{\y}\WnvecXm^{\frac{1}{1+\rho}}\bigg(\sum_{m'\neq m}\WnvecXmp^{\frac{1}{1+\rho}}\bigg)^{\rho}\Bigg)^s\right] \\
%&=  \sum_{m=1}^{M_n}\EE\left[\Bigg(\frac{1}{M_n}\sum_{\y}\WnvecXm^{\frac{1}{1+\rho}}\bigg(\sum_{m'\neq m}\WnvecXmp^{\frac{1}{1+\rho}}\bigg)^{\rho}\Bigg)^s\right] \\\label{eqn:psum_bound_rce_s_mean_1_a}
& =  \sum_{m=1}^{M_n}\EE\left[\Bigg(\frac{1}{M_n}\sum_{\y}\WnvecXm^{\frac{1}{1+\rho}}\bigg(\sum_{m'=1}^{M_n}\WnvecXmp^{\frac{1}{1+\rho}}\bigg)^{\rho}\Bigg)^s\right] \\\label{eqn:psum_bound_rce_s_mean_1}
&=  {M_n}\EE\left[\Bigg(\frac{1}{M_n}\sum_{\y}\WnvecX^{\frac{1}{1+\rho}}\bigg(\sum_{m'=1}^{M_n}\WnvecXmp^{\frac{1}{1+\rho}}\bigg)^{\rho}\Bigg)^s\right],
\end{align}
where \eqref{eqn:psum_bound_rce_s_mean_1} is because $m$ is a dummy variable and, after averaging, it leads to $M_n$ equal terms. Rising \eqref{eqn:psum_bound_rce_s_mean_1} to $\frac{1}{s}$, we obtain the following bound on the tilted average probability of error:
\begin{align}\label{eqn:psum_bound_rce_s_mean_2}
\mathbb{E}[\Pecn^s]^{\frac{1}{s}} \leq  \frac{1}{M_n}M_n^{\frac{1}{s}}\Bigg(\EE\left[\Bigg(\sum_{\y}\WnvecX^{\frac{1}{1+\rho}}\bigg(\sum_{m'=1}^{M_n}\WnvecXmp^{\frac{1}{1+\rho}}\bigg)^{\rho}\Bigg)^s\right]\Bigg)^{\frac{1}{s}}.
\end{align}
Now we proceed to upper bound the right-hand side of \eqref{eq:5} in two different ways starting from \eqref{eqn:psum_bound_rce_s_mean_2}. We show that one of the two bounds is $E_{{\rm ex}}^n(2R,Q^n) +R$, where $\Eexn$ is Gallager's expurgated exponent, while the other is the random-coding exponent. In Remark~\ref{rem:4} after Theorem~\ref{theo:2} we show that, under certain conditions, taking the maximum of the two and optimizing over the input distribution we recover the \ac{TRC} exponent for the whole range of rates below capacity.

\paragraph{Derivation of $\Etrclbl$}\label{sec:Etrclbl}
Restricting $\rho$ to $[0,1]$ in \eqref{eqn:psum_bound_rce_s_mean_2} while keeping $s\in(0,1]$ we have the following set of steps
\begin{align}\label{eqn:trc_from_gallager}
\mathbb{E}[\Pecn^s]^{\frac{1}{s}}  &\leq \frac{1}{M_n}M_n^{\frac{1}{s}}\Bigg(\sum_{\x}Q^n(\x)\EE\left[\Bigg(\sum_{\y}\Wnvec^{\frac{1}{1+\rho}}\bigg(\sum_{m'=1}^{M_n}\WnvecXmp^{\frac{1}{1+\rho}}\bigg)^{\rho}\Bigg)^s\middle|\X=\x\right]\Bigg)^{\frac{1}{s}}  \\\label{eqn:trc_from_gallager_2}
&\leq  \frac{1}{M_n}M_n^{\frac{1}{s}}\Bigg(\sum_{\x}Q^n(\x)\EE\left[\Bigg(\sum_{\y}\Wnvec^{\frac{1}{1+\rho}}\sum_{m'=1}^{M_n}\WnvecXmp^{\frac{\rho}{1+\rho}}\Bigg)^s\middle|\X=\x\right]\Bigg)^{\frac{1}{s}}  \\\label{eqn:trc_from_gallager_3}
&\leq  \frac{1}{M_n}M_n^{\frac{1}{s}}\Bigg(\sum_{\x}Q^n(\x)\EE\left[\sum_{m'=1}^{M_n}\Bigg(\sum_{\y}\Wnvec^{\frac{1}{1+\rho}}\WnvecXmp^{\frac{\rho}{1+\rho}}\Bigg)^s\middle|\X=\x\right]\Bigg)^{\frac{1}{s}} \\\label{eqn:trc_from_gallager_4}
&= \frac{1}{M_n}M_n^{\lambda}\Bigg(\sum_{\x}Q^n(\x)\sum_{m'=1}^{M_n}\EE\left[\Bigg(\sum_{\y}\Wnvec^{1-t}\WnvecXmp^{t}\Bigg)^{\frac{1}{\lambda}}\middle|\X=\x\right]\Bigg)^{\lambda}  \\
&= \frac{1}{M_n}\big(M_n^2\big)^{\lambda}\Bigg(\sum_{\x}\sum_{\x'}Q^n(\x)Q^n(\x'|\x)\Bigg(\sum_{\y}\Wnvec^{1-t}\Wnvecp^{t}\Bigg)^{\frac{1}{\lambda}}\Bigg)^{\lambda}  \\
&= \frac{1}{M_n}\big(M_n^2\big)^{\lambda}\Bigg(\sum_{\x}\sum_{\x'}Q^n(\x)Q^n(\x')\Bigg(\sum_{\y}\Wnvec^{1-t}\Wnvecp^{t}\Bigg)^{\frac{1}{\lambda}}\Bigg)^{\lambda} \label{eqn:trc_from_gallager_last}\\
&\leq\frac{1}{M_n}\big(M_n^2\big)^{\lambda}\Bigg(\sum_{\x}\sum_{\x'}Q^n(\x)Q^n(\x')\Bigg(\sum_{\y}\sqrt{\Wnvec\Wnvecp}\Bigg)^{\frac{1}{\lambda}}\Bigg)^{\lambda} \label{eqn:trc_from_gallager_last_+}
\end{align}
where in \eqref{eqn:trc_from_gallager} we conditioned the expectation to a given $\x$ and averaged out over all possible $\x$, \eqref{eqn:trc_from_gallager_2} follows from the inequality $(\sum_i a_i)^{\rho}\leq \sum_i (a_i)^{\rho}$ for $0\leq\rho \leq 1$, \eqref{eqn:trc_from_gallager_3} follows from the same inequality considered with respect to $s$ instead of $\rho$, 
equality \eqref{eqn:trc_from_gallager_4} comes from renaming the variables $s\rightarrow 1/\lambda$, $\rho\rightarrow t/(1-t)$ (to make the expression aesthetically similar to Gallager's expurgated), changing their ranges to $\lambda\in[1,\infty)$, $t\in[0,1/2]$ and from the linearity of the expectation, while \eqref{eqn:trc_from_gallager_last} is because $m'$ is a dummy variable of summation leading to $M_n-1$ identical terms and from the fact that we consider pairwise-independent codewords, which implies $Q^n(\x'|\x)=Q^n(\x')$, where $Q^n(\x'|\x)$ is the probability to generate codeword $\x'$ given that codeword $\x$ has been already generated. Now we note that \eqref{eqn:trc_from_gallager_last} is convex in $t$ %\cite[Expression (5.5.2)]{gallagerBook} 
and symmetric with respect to $t=1/2$. Thus, \eqref{eqn:trc_from_gallager_last} is minimized with respect to $t$ in $t=1/2$,  leading to expression \eqref{eqn:trc_from_gallager_last_+}.
Taking the negative normalized logarithm of \eqref{eqn:trc_from_gallager_last_+}, we obtain for $s\in(0,1]$ and $\lambda\in[1,\infty)$ that
\begin{align}\label{eqn:exp_Gall_bound_lr}
-\frac{1}{n}\log\mathbb{E}[\Pecn^s]^{\frac{1}{s}} \geq  R -\lambda2R -\frac{1}{n}\lambda\log\Bigg(\sum_{\x}\sum_{\x'}Q^n(\x)Q^n(\x')\Bigg(\sum_{\y}\sqrt{\Wnvec\Wnvecp}\Bigg)^{\frac{1}{\lambda}}\Bigg),
\end{align}
where we used the fact that $\log M_n = nR$. Including the $\gamma_n$ in \eqref{eqn:exp_Gall_bound_lr} and optimizing both sides in the respective variables, we have
\begin{align}\label{eqn:exp_Gall_bound_lr_opt}
\max_{s\in(0,1]} -\frac{1}{s}\frac{1}{n} \log \gamma_n -\frac{1}{n}\log\mathbb{E}[\Pecn^s]^{\frac{1}{s}} \geq &  \max_{\lambda\in[1,\infty)}-\frac{\lambda}{n} \log \gamma_n + R -\lambda2R - ~\notag\\&\frac{1}{n}\lambda\log\Bigg(\sum_{\x}\sum_{\x'}Q^n(\x)Q^n(\x')\Bigg(\sum_{\y}\sqrt{\Wnvec\Wnvecp}\Bigg)^{\frac{1}{\lambda}}\Bigg).
\end{align}

\paragraph{Derivation of $\Etrclbh$}\label{sec:Etrclbh}
In the following we derive the random-coding exponent starting from \eqref{eqn:psum_bound_rce_s_mean_1}. 
Restricting $\rho$ to $[0,1]$ in \eqref{eqn:psum_bound_rce_s_mean_2} while keeping $s\in(0,1]$ we have now that
\begin{align}\label{eqn:trc_Gall_bound_1}
\mathbb{E}[\Pecns^s]^{\frac{1}{s}}  &\leq  \frac{1}{M_n}M_n^{\frac{1}{s}}\Bigg(\sum_{\x}Q^n(\x)\EE\left[\Bigg(\sum_{\y}\Wnvec^{\frac{1}{1+\rho}}\bigg(\sum_{m'=1}^{M_n}\WnvecXmp^{\frac{1}{1+\rho}}\bigg)^{\rho}\Bigg)^s\middle|\X=\x\right]\Bigg)^{\frac{1}{s}}\\\label{eqn:trc_Gall_bound_2}
&\leq \frac{1}{M_n}M_n^{\frac{1}{s}}\Bigg(\sum_{\x}Q^n(\x)\Bigg(\sum_{\y}\Wnvec^{\frac{1}{1+\rho}}\EE\left[\bigg(\sum_{m'=1}^{M_n}\WnvecXmp^{\frac{1}{1+\rho}}\bigg)^{\rho}\middle|\X=\x\right]\Bigg)^s\Bigg)^{\frac{1}{s}}  \\\label{eqn:trc_Gall_bound_3}
&\leq \frac{1}{M_n}M_n^{\frac{1}{s}}\Bigg(\sum_{\x}Q^n(\x)\Bigg(\sum_{\y}\Wnvec^{\frac{1}{1+\rho}}\bigg(\sum_{m'=1}^{M_n}\EE\left[\WnvecXmp^{\frac{1}{1+\rho}}\middle|\X=\x\right]\bigg)^{\rho}\Bigg)^s\Bigg)^{\frac{1}{s}} \\
&= \frac{1}{M_n}M_n^{\frac{1}{s}}M_n^{\rho}\Bigg(\sum_{\x}Q^n(\x)\Bigg(\sum_{\y}\Wnvec^{\frac{1}{1+\rho}}\bigg(\sum_{\x'}Q^n(\x'|\x)\Wnvecp^{\frac{1}{1+\rho}}\bigg)^{\rho}\Bigg)^s\Bigg)^{\frac{1}{s}}  \label{eqn:trc_Gall_bound_final}\\
&= \frac{1}{M_n}M_n^{\frac{1}{s}}M_n^{\rho}\Bigg(\sum_{\x}Q^n(\x)\Bigg(\sum_{\y}\Wnvec^{\frac{1}{1+\rho}}\bigg(\sum_{\x'}Q^n(\x')\Wnvecp^{\frac{1}{1+\rho}}\bigg)^{\rho}\Bigg)^s\Bigg)^{\frac{1}{s}}  \label{eqn:trc_Gall_bound_final_new}\\
&= \frac{1}{M_n}M_n^{t}M_n^{\rho}\Bigg(\sum_{\x}Q^n(\x)\Bigg(\sum_{\y}\Wnvec^{\frac{1}{1+\rho}}\bigg(\sum_{\x'}Q^n(\x')\Wnvecp^{\frac{1}{1+\rho}}\bigg)^{\rho}\Bigg)^{\frac{1}{t}}\Bigg)^{t} \label{eqn:trc_Gall_bound_final_varrho}\\
&\leq \frac{1}{M_n}M_n^{t}M_n^{\rho}\Bigg(\Bigg(\sum_{\x}Q^n(\x)\sum_{\y}\Wnvec^{\frac{1}{1+\rho}}\bigg(\sum_{\x'}Q^n(\x')\Wnvecp^{\frac{1}{1+\rho}}\bigg)^{\rho}\Bigg)^{\frac{1}{t}}\Bigg)^{t}  \label{eqn:trc_Gall_bound_final_varrho_1}\\
&\leq \frac{1}{M_n}M_n^{t}M_n^{\rho}\Bigg(\sum_{\x}Q^n(\x)\sum_{\y}\Wnvec^{\frac{1}{1+\rho}}\bigg(\sum_{\x'}Q^n(\x')\Wnvecp^{\frac{1}{1+\rho}}\bigg)^{\rho}\Bigg)  \label{eqn:trc_Gall_bound_final_varrho_2}\\
&\leq \frac{1}{M_n}M_n^{t}M_n^{\rho}\Bigg(\sum_{\y}\bigg(\sum_{\x}Q^n(\x)\Wnvecp^{\frac{1}{1+\rho}}\bigg)^{1+\rho}\Bigg)  ,\label{eqn:trc_Gall_bound_final_varrho_3}
\end{align}
where \eqref{eqn:trc_Gall_bound_2} follows from Jensen's inequality and the concavity of $x^s$ for $s\in(0,1]$, \eqref{eqn:trc_Gall_bound_3}  follows from Jensen's inequality and the concavity of $x^{\rho}$ for $\rho\in[0,1]$, \eqref{eqn:trc_Gall_bound_final} comes from the fact that $m'$ is a dummy variable once the average over $\x'$ is taken leading to $\left(\left(M_n^{\rho}\right)^s\right)^{1/s}=M_n^{\rho}$ equal terms, \eqref{eqn:trc_Gall_bound_final_new} is because of pairwise independence of the codewords, in \eqref{eqn:trc_Gall_bound_final_varrho} we applied the change of variable $s\rightarrow 1/t$, $t\in[1,\infty)$, while \eqref{eqn:trc_Gall_bound_final_varrho_1} is because of Jensen's inequality applied to the function $x^\frac{1}{t}$ which is concave for $t \geq 1$.
Taking the negative normalized logarithm of \eqref{eqn:trc_Gall_bound_final_varrho_3}, including $\gamma_n$ and optimizing we find
\begin{align}
\max_{s\in(0,1]}  -\frac{1}{s}\frac{1}{n}\log \gamma_n -\frac{1}{n}\log \mathbb{E}[\Pecn^s]^{\frac{1}{s}}  & \geq \max_{t\geq 1}\max_{\rho\in [0,1]} -\frac{t}{n}\log \gamma_n  -\rho R  -(t-1)R\notag
\\& \hspace{7em}-\frac{1}{n}\log\Bigg(\sum_{\y}\bigg(\sum_{\x}Q^n(\x)\Wnvecp^{\frac{1}{1+\rho}}\bigg)^{1+\rho}\Bigg)\notag\\\label{eqn:exp_Gall_bound_hr}
&\geq \max_{\rho\in [0,1]} -\frac{1}{n}\log \gamma_n -\rho R -\frac{1}{n}\log\Bigg(\sum_{\y}\bigg(\sum_{\x}Q^n(\x)\Wnvecp^{\frac{1}{1+\rho}}\bigg)^{1+\rho}\Bigg),
\end{align}
where \eqref{eqn:exp_Gall_bound_hr} is because  $t=1$ leads to the maximum.

Combining \eqref{eqn:exp_Gall_bound_lr_opt} and \eqref{eqn:exp_Gall_bound_hr} and using Lemma \ref{lem:main}  we obtain the theorem statement.
\end{IEEEproof}

Theorem \ref{theo5}, since based on Lemma~\ref{lem:main}, only requires $\gamn$ to take positive values. By adding some additional constraint on $\gamn$, the next theorem gives a bound on the exponents of typical codes. Like Theorem \ref{theo5}, the following theorem holds for channels with arbitrary alphabets or memory and pairwise-independent code ensembles. 

\begin{theorem}\label{theo:2}
Let $\gamn$ be a positive real-valued sequence in $n$ such that 
\begin{gather}\label{eqn:gamm_theo_2}
\liminf_{n\rightarrow\infty}\gamn=\infty,\\
\sum_{n=1}^\infty\frac{1}{\gamn}<\infty.
\end{gather}
Then, for any channel $W^n$ and pairwise-independent ensemble with codeword distribution $Q^n$ it holds that:
\begin{equation}\label{eqn:theo5_statement2}
%\PP\Bigl [\liminf_{n\rightarrow\infty} \En  > \liminf_{n\rightarrow\infty}	\min\left\{ \Etrclbl,\Etrclbh\right\} \Bigr ]=1.
\PP\Bigl [\liminf_{n\rightarrow\infty} \En  > \liminf_{n\rightarrow\infty}	\Etrclbmax\Bigr ]=1.
\end{equation} 
\end{theorem}
\begin{IEEEproof}
The proof is based on the Borel-Cantelli lemma. Such lemma is often used to upgrade convergence in probability to almost sure (a.s.) convergence. However, our theorem does not consider either kind of convergence, although the result of the theorem is similar in spirit to going from convergence in probability to a.s.~convergence.

From the assumptions and from the statement of Theorem \ref{theo5} we have that $
 \lim_{n\rightarrow\infty}\PP\bigl [\mathcal{A}_n(\Cn) ]=1$ where
\begin{align}\label{eqn:eventAn}
\mathcal{A}_n(\Cns) = \left\{\Ens  \leq	\Etrclbmax\right\}.
\end{align}
Consider also the following asymptotic event \cite[Sec.~1.3]{probPath_Resnick1998}\cite[Sec.~2.3]{Durett_probab_Book}:
\begin{align}\label{eqn:eventA}
\limsup_{n\rightarrow\infty} \mathcal{A}_n(\mathcal{C}_n) = \bigcap_{n\geq 1}\bigcup_{k\geq n}\mathcal{A}_k(\mathcal{C}_k) = \{\mathcal{A}_n(\Cns)\text{ i.o.}\},
\end{align}
where ``i.o.'' stands for ``infinitely often'' and denotes that events of the kind $\mathcal{A}_n(\Cns)$ happen for an infinite number of values of $n$. The event in \eqref{eqn:eventAn} corresponds to finding a code for which the event within square brackets in \eqref{eqn:theo5_statement} does not hold (i.e.~the complementary event).
The event in \eqref{eqn:eventA} corresponds to the event ``infinitely many $\mathcal{A}_n(\Cns)$ occur'' . Note that in \eqref{eqn:eventA} the code sequence $\Cns, n=1,\ldots,\infty$, is fixed.

To get an intuition on how this corresponds to the left-hand side term in \eqref{eqn:eventA}, assume there exists a positive integer $n'$ such that $\bigcup_{k\geq n'}\mathcal{A}_k(\mathcal{C}_k)=\emptyset$, i.e., no $\mathcal{A}_k(\mathcal{C}_k)$ occurs for $k\geq n'$. In this case less than $n'$ events of the kind $\mathcal{A}_n(\Cns)$ can occur, and so only finitely many such events happen. Note also the following: consider the event $\bigcap_{n\geq 1}\mathcal{A}_n(\Cns)$. While the occurrence of such event implies that infinitely many $\mathcal{A}_n(\Cns)$ occur, it also implies that each $\mathcal{A}_n(\Cns), n=1,2,\ldots$ occur, which is not needed here. Hence the union in \eqref{eqn:eventA}.  From Theorem \ref{theo5} and from \eqref{eqn:gamm_theo_2}, we have that $
\PP\bigl[ \mathcal{A}_n(\Cn)\bigr]\leq \frac{1}{\gamma_n}$, and as a consequence that
\begin{align}\label{eqn:Borel_condition}
\sum_{n=1}^{\infty}\PP\Bigl[ \mathcal{A}_n(\Cn)\Bigr]< \infty.
\end{align}
 By the Borel-Cantelli lemma \cite[Sec.~2.3]{Durett_probab_Book} and using \eqref{eqn:Borel_condition} we have:
\begin{align}\label{eqn:Borel_0}
\PP\Bigl[\mathcal{A}_n(\Cn)\text{ i.o.}\Bigr]=0.
\end{align}
Equation~\eqref{eqn:Borel_0} implies that the event complementary to the one within brackets has probability one, that is
\begin{align}\label{eqn:Borel_2}
\PP\Bigl[\overline{\left\{\mathcal{A}_n(\Cn)\text{ i.o.}\right\}}\Biggr]&=\PP\Biggl[\overline{\biggl\{\limsup_{n\rightarrow\infty} \mathcal{A}_n(\Cn)\biggr\}}\Biggr]\\\label{eqn:Borel_2_1}
&=\PP\left[\liminf_{n\rightarrow\infty}\overline{\mathcal{A}_k(\mathsf{C}_k)}\right]\\\label{eqn:Borel_2_2}
&=1,
\end{align}
where to obtain \eqref{eqn:Borel_2_1} we applied twice the De Morgan's law \cite[Sec.~1.2]{probPath_Resnick1998}.
From \eqref{eqn:Borel_2_2} we obtain the theorem statement.
\end{IEEEproof}

Some considerations are in order. 
\begin{remark}
We use $\liminf_{n\rightarrow\infty}$ rather than $\lim_{n\rightarrow\infty}$ in \eqref{eqn:theo5_statement2} because on the one hand $\liminf_{n\rightarrow\infty} \En $ necessarily exists in $\mathbb{R}\cup\{-\infty,+\infty\}$. %since $\liminf_{n\rightarrow\infty}$ is increasing in $n$
 On the other hand, for a given code the limit of the exponent might not exist even if the limit of $E_{{\rm ex}}^n(2R,Q^n)$ does.
Note that we used the fact that $R=\liminf_{n\rightarrow\infty}\frac{1}{n}\log M_n=\limsup_{n\rightarrow\infty}\frac{1}{n}\log M_n$, which follows from the definition of $R$.
\end{remark}

\begin{remark}\label{rem:3}
Theorem \ref{theo:2} holds whether the positive sequence $\delta_n$ has a limit or not and, if it does, whether it converges or not. Let us consider the case in which it converges, that is
\begin{align}\label{eqn:delta_conv}
\delta_n=\frac{\hat\rho_n}{n} \log \gamma_n \to  c,
\end{align}
where $c<\infty$. The smaller $c$ is, the tighter the bound. Hence, we are particularly interested in the case $c=0$. Note that \eqref{eqn:delta_conv} implies $\iota_n \to 0$, since $\iota_n\leq \delta_n$.
Define $\hat \rho = \lim_{n\to\infty} \hat\rho_n$. Assume that $\hat\rho<\infty$. Then we need that $\lim_{n\to\infty}\delta_n = \lim_{n\to\infty}\frac 1n \log \gamma_n=0$, which imposes a further constraint on $\gamma_n$. Instead, for $\hat\rho_n\to\infty$ we have that $\lim_{n\to\infty}\delta_n = 0$ if and only if $\hat\rho_n$ grows slower than $\frac{n}{\log \gamma_n}$, e.g., as  $\hat\rho_n=\frac{\sqrt{n}}{\log\gamn}$. Under these assumptions, $\delta_n$ vanishes with $n$.
Thus, $\gamma_n$ must grow fast enough for the series of its reciprocal to converge, but slow enough so that the above holds. 
This is the case, for example, for i.i.d. codes and constant-composition codes over \ac{DMC} as well as any code over finite-state channels, as we show in the next section. 
We analyze the case in which $\hat\rho_n\to\infty$ in Section~\ref{sec:unbounded_rho}.
\end{remark}

In some case it can happen that, although $\iota_n$ and $\delta_n$ go to $0$ asymptotically, the statement of Lemma \ref{lem:main} has limited practical relevance, despite the conditions of Theorem \ref{theo:2} being satisfied. This is the case, for instance, in channels whose capacity is zero. As an example, we consider the quasi-static \ac{BSC} studied in \cite{font2017}, where the crossover probability of the channel $p$ is fixed for the whole transmission of a codeword, and changes randomly according to some distribution, from codeword to codeword. In this case, the ensemble-average error probability of the quasi-static \ac{BSC} does not vanish with $n$ but rather converges to a constant, the outage probability, given by $P_{\rm out}(R)=\mathbb{P}[I(p)<R]$, where $I(p)=1-h(p)$ is the mutual information of a BSC with crossover probability $p$, where the probability is taken with respect to the random crossover probability $p$. In this case, the complementary of the tail probability in~\eqref{eqn:theo5_statement2}, written in terms of the error probability instead of the error exponent, with $s=1$ for simplicity, satisfies
\begin{align}\label{eq:4_example}
\mathbb{P}\left[ \liminf_{n\rightarrow\infty}\Pecn =  \liminf_{n\rightarrow\infty}\gamma_n \, \mathbb{E}[P_{\rm e}(\Cn,p)] \right]&  = \mathbb{P}\left[ \liminf_{n\rightarrow\infty}\Pecn \geq \liminf_{n\rightarrow\infty} \gamma_n \cdot P_{\rm out}(R) \right]\\
&= 
\mathbb{P}\left[ \liminf_{n\rightarrow\infty}\Pecn \geq  \infty \right]\\
&= 0,
\end{align}
where we assumed that $P_{\rm out}(R)>0$, in other words, that the distribution of the crossover probability $p$ is such that the expectation in the left-hand side of~\eqref{eq:4_example} converges to a positive number. For such channel, Theorem \ref{theo:2} applies but gives a result which is not practically relevant.

There might also be cases that $\delta_n$ diverges for $R=0$. Whether for a given pairwise code ensemble and channel this is the case or not, depends on the variance of the quantity $\log Z_n(\x,\x')$, as will be shown in Section \ref{sec:unbounded_rho}.

\begin{remark}\label{rem:4}
Let $E_{{\rm ex}}(R,Q)$ and $\Er$ be the asymptotic single-letter expurgated and random-coding exponents, respectively in~\eqref{eqn:expu} and~\eqref{eqn:rce}, and denote by $\Q$ the limiting distribution of $Q^n$. If the following limits exist
\begin{gather}\label{eqn:theo5_limit}
\lim_{n\to\infty} E_{{\rm ex}}^n(2R,Q^n) = E_{{\rm ex}}(2R,Q),\\
\label{eqn:theo5_limit2}
\lim_{n\to\infty} \Ern = \Erbold,
\end{gather}
after exploiting the superadditivity of $\liminf$ in Theorem \ref{theo:2}, it follows that
\begin{equation}\label{eqn:theo5_statement2B}
\PP \left[\liminf_{n\rightarrow\infty} \En  > \max\left\{E_{\rm ex}(2R,\Q)+R-\liminf_{n\rightarrow\infty}\delta_n , \Erbold-\liminf_{n\rightarrow\infty}\iota_n\right\}\right]=1.
\end{equation} 
Equation~\eqref{eqn:theo5_statement2B} holds independently on the behaviour of $\delta_n$ and $\iota_n$ and specifically on whether they (or only one of them) approaches $0$ as $n$ tends to infinity. Note that if they both diverge, \eqref{eqn:theo5_statement2B} leads to a trivial result while if only $\delta_n$ tends to infinity while $\iota_n\rightarrow 0$ the maximum always coincides with $\Erbold$.

Restricting to channels and ensembles for which $\delta_n \to 0$ (and thus $\iota_n \to 0$, see Remark \ref{rem:3}), Equation~\eqref{eqn:theo5_statement2B} becomes
\begin{equation}\label{eqn:theo5_statement2_conv}
\PP \left[\liminf_{n\rightarrow\infty} \En  > \max\left\{E_{\rm ex}(2R,\Q)+R, \Erbold\right\}\right]=1.
\end{equation} 
Now, keeping the assumption that the limits exist and that \eqref{eqn:theo5_statement2_conv} holds, consider the maximum $R$ (if it exists) such that the optimal $\rho$ leading to $\Er$ is $1$ and let us refer to such rate as $R^*$. We have: $E_{\rm ex}(R^*,\Q)= E_{\rm r}(R^*,\Q)$, which follows from similar arguments as in \cite[Sec.~5.7]{gallagerBook}. Furthermore, since the derivatives of $E_{\rm ex}(R,\Q)$ and $\Erbold$ with respect to $R$ are exactly $-\lambda$ and $-\rho$, respectively, we obtain that
\begin{equation}\label{eqn:theo_lim_finite_sta1_gen}
\max\left\{E_{\rm ex}(2R,\Q)+R, \Er\right\} =
\begin{cases}
E_{\rm ex}(2R,\Q)+R  &  R\leq R^* \\
E_{\rm r}(R,\Q)  &  R > R^*,
\end{cases}
\end{equation}
recovering the \ac{TRC} exponent over all rates. We note that $E_{\rm r}(R,\Q)$ might be zero above a certain rate. Optimizing over the input distribution $\Q$, such rate indeed coincides with the channel capacity. Note also that, in case $\delta_n\rightarrow\infty$ and $\iota_n\rightarrow 0$, then $R^*=0$, i.e., $\Etrclbmax$ coincides with $E_{\rm r}(R,\Q)$.
\end{remark}

Observe that Theorem~\ref{theo5} does not apply to code ensembles whose codeword generation is not pairwise independent, such as the random Gilbert-Varshamov (RGV) ensemble \cite{generalizedGV_fabregas2020}. For this ensemble, it is known that the exponent of the ensemble average error probability is the maximum of the expurgated and random-coding exponents \cite{generalizedGV_fabregas2020}. For ensembles whose average error probability achieves the expurgated exponent, like the RGV, by setting $s=1$ in Lemma \ref{lem:main}, we find that $
\PP \Big[ \En > \Eexn - \delta_n\Big]>1-\epsilon_n$,  implying that the error exponent of typical codes is lower bounded by $\Eexn$. This can also be seen as a consequence of Jensen's inequality, since  $\EE[\log P_{\rm e}(\Cn)]\leq \log\EE[P_{\rm e}(\Cn)]=\Eexn$, and thus
 $\Etrcbold = \lim_{n\to\infty}-\frac{1}{n}\EE[\log P_{\rm e}(\Cn)]\geq\Eexn$.

%%%%%%%%%%%%%%%%%%%%%%%%%%
\subsection{Mismatched Decoder}\label{sec:mismatched}
While Theorem \ref{theo5} and  Theorem \ref{theo:2} assume a \ac{ML} decoder, they can be easily generalized to mismatched decoding, that recovers \ac{ML} as a special case. This is done by replacing $\Etrclbl$ and $\Etrclbh$ with their mismatched counterpart as follows. Let $q^n(\x,\y)$ be the decoding metric, $\mathcal{D}_m$ the decision region for codeword $\x_m$ according to such metric and $\hat{m}$  the index of the codeword selected by the decoder. The mismatched decoder uses the decoding rule
\begin{align}\label{eqn:md_rule}
\hat{m}=\argmax_{m\in\{1,\ldots,M_n\}} q^n(\x_m,\y).
\end{align}
The choice $q^n(\x,\y)=\Wnvecm$ recovers the \ac{ML} decoder.
If codeword $\x_m$ is transmitted, for all $\y\notin \mathcal{D}_m$, where $ \mathcal{D}_m$ is the decoding region for message $m$, we have that
\begin{align}\label{eqn:md_error_event}
\Bigg( \sum_{m'\neq m}\bigg( \frac{q^n(\x_{m'},\y)}{q^n(\x_{m},\y)} \bigg)^\tau \Bigg)^\rho\geq 1,
\end{align}
for $\tau \geq 0,\rho\geq 0$. Using \eqref{eqn:md_error_event}, we can bound the probability of error $\Pecnm$ and find  an upper bound on the probability of error for a given code in the ensemble. Similarly to \eqref{eqn:psum_bound_rce}, we have
\begin{align}\label{eqn:psum_bound_rce_md3}
\Pecns %& =  \frac{1}{M_n}\sum_{m=1}^{M_n}\Pecnm\\\label{eqn:psum_bound_rce_md2}
%&= \frac{1}{M_n}\sum_{m=1}^{M_n}\sum_{\y\notin \mathcal{D}_m}\Wnvecm\cdot 1\\
&\leq \frac{1}{M_n}\sum_{m=1}^{M_n}\sum_{\y}\Wnvecm \Bigg( \sum_{m'\neq m}\bigg( \frac{q^n(\x_{m'},\y)}{q^n(\x_{m},\y)} \bigg)^{\tau} \Bigg)^\rho, 
\end{align}
where \eqref{eqn:psum_bound_rce_md3} follows from \eqref{eqn:md_error_event} and from extending the sum to all $\y$. 

Using \eqref{eqn:psum_bound_rce_md3} instead of \eqref{eqn:psum_bound_rce} in the proof of Theorem~\ref{theo5} we obtain that $\Etrclbh$ and $\Etrclbl$ are now respectively given by
\begin{align}\label{eqn:EtrclbhMD}
\Etrclbh&= \max_{\rho\in [0,1],\tau\geq 0} -\frac{1}{n}\log \gamma_n -\rho R -\frac{1}{n}\log\Bigg(\sum_{\x}Q^n(\x)\sum_{\y}\Wnvec\bigg(\sum_{\x'}Q^n(\x')\left(\frac{q^n(\x',\y)}{q^n(\x,\y)}\right)^{\tau}\bigg)^{\rho}\Bigg) 
\end{align}
and
\begin{align}\label{eqn:EtrclblMD}
\Etrclbl = \max_{\lambda\geq 1, \tau\geq 0}-\frac{\lambda}{n} \log \gamma_n + R -\lambda2R -\frac{\lambda}{n}\log\Bigg(\sum_{\x}\sum_{\x'}Q^n(\x)Q^n(\x')\Bigg(\sum_{\y}\Wnvec\bigg(\frac{q^n(\x',\y)}{q^n(\x,\y)}\bigg)^\tau\Bigg)^{\frac{1}{\lambda}}\Bigg).
\end{align}
Note that, except for the term $-\frac{1}{n}\log \gamma_n $, the result in~\eqref{eqn:EtrclbhMD} recovers the expression of Gallager's  mismatched decoding exponent for i.i.d. inputs~\cite[Eq.~(7.4)]{mismathedBookScarlett2020}. Similarly, subtracting $R$ and $-\frac{\lambda}{n} \log \gamma_n$ from \eqref{eqn:EtrclblMD} and substituting $2R$ with $R$ in the remaining expression, one recovers the expurgated exponent for the mismatched decoding i.i.d. case given in \cite[Eq.~(7.35)]{mismathedBookScarlett2020}. Using \eqref{eqn:EtrclbhMD} and \eqref{eqn:EtrclblMD}, Theorem \ref{theo5} and Theorem \ref{theo:2} can be directly extended to the mismatched case.

\subsection{Low Rates, $\hat\rho_n\to\infty$}\label{sec:unbounded_rho}
For some channels and ensembles it can happen that, in the rate regime where $\Etrclbmax=\Etrclbl$, the optimal $\lambda$ grows unbounded with $n$, i.e., $\hat\lambda_n\to\infty$. This is easy to see for $R=0$, but can also happen for positive rates. We start by calculating the derivative of $\Exnvarrho$ with respect to $\lambda$ and show that it is strictly positive and, except some special cases, for $n\geq 2$ and $1\leq \lambda<\infty$. We have: 
\begin{align}\label{eqn:deriv_F}
\frac{\partial \Exnvarrho}{\partial \lambda}=&-\frac{1}{n}\log\sum_{\x}\sum_{\x'} Q^n(\x)Q^n(\x')Z^n(\x,\x')^{\frac{1}{\lambda}}  +\frac{1}{n}\frac{\sum_{\x}\sum_{\x'} Q^n(\x)Q^n(\x')\log \bigl( Z^n(\x,\x')^{\frac{1}{\lambda}}\bigr)Z^n(\x,\x')^{\frac{1}{\lambda}}}{\sum_{\x}\sum_{\x'} Q^n(\x)Q^n(\x')Z^n(\x,\x')^{\frac{1}{\lambda}}}
\\\label{eqn:deriv_F1}
=& -\frac{1}{n}\log\mathbb{E}_{Q^n}\left[Z^n(\X,\X')^{\frac{1}{\lambda}}\right]  + \frac{1}{n}\frac{\sum_{\x}\sum_{\x'} Q^n(\x)Q^n(\x')Z^n(\x,\x')^{\frac{1}{\lambda}}\log\left[Z^n(\x,\x')^{\frac{1}{{\lambda}}}\right]}{\mathbb{E}_{Q^n}\left[Z^n(\X,\X')^{\frac{1}{\lambda}}\right]}
\\\label{eqn:deriv2}
=& -\frac{1}{n}\log\mathbb{E}_{Q^n}\left[Z^n(\X,\X')^{\frac{1}{\lambda}}\right]  + \frac{1}{n}\sum_{\x}\sum_{\x'} \Pjoint^n(\x,\x')\log\frac{\Pjoint^n(\x,\x')}{Q^n(\x)Q^n(\x')} + ~ \notag\\ &~\hspace{2em}~+\frac{1}{n}\sum_{\x}\sum_{\x'}\Pjoint^n(\x,\x')\log\left\{\mathbb{E}_{Q^n}\left[Z^n(\X,\X')^{\frac{1}{\lambda}}\right] \right\}\\\label{eqn:deriv2_1}
=&  \frac{1}{n}\sum_{\x}\sum_{\x'} \Pjoint^n(\x,\x')\log\frac{\Pjoint^n(\x,\x')}{Q^n(\x)Q^n(\x')}
\\\label{eqn:deriv3}
=& \frac{1}{n}D\left({\Pjoint^n}\|Q^nQ^n\right)
\end{align}
where in \eqref{eqn:deriv_F1} we substituted
\begin{align}\label{eqn:deriv3_1}
\mathbb{E}_{Q^n}\left[Z(\X,\X')^{\frac{1}{\lambda}}\right] =   \sum_{\x}\sum_{\x'} Q^n(\x)Q^n(\x')Z^n(\x,\x')^{\frac{1}{\lambda}}.
\end{align}
In \eqref{eqn:deriv2} we defined
\begin{align}\label{eqn:psi_def}
\Pjoint^n(\x,\x') = \frac{Q^n(\x)Q^n(\x')Z^n(\x,\x')^{\frac{1}{\lambda}}}{\mathbb{E}_{Q^n}\left[Z(\X,\X')^{\frac{1}{\lambda}}\right]},
\end{align}
while in \eqref{eqn:deriv2_1} we used the fact that ${\Pjoint}^n(\x,\x')$ is a probability distribution and thus its components add up to $1$.
As suggested by Equation~\eqref{eqn:deriv3}, the derivative of $E_{\rm x}^{n}(\lambda,Q^n)$ is $\frac 1n$ times the relative entropy between the distributions ${\Pjoint^n(\x,\x')}$ and $Q^n(\x)Q^n(\x')$, and thus is strictly positive except in some special case. % $Z(\x,\x')^{\frac{1}{\lambda}}=1$ for all pairs $(\x,\x')$, in which case takes value $0$. 
Furthermore, for most channels and ensembles, for any finite $\lambda\geq 1$ its value is bounded. This is because the relative entropy takes value $\infty$ only if the ratio of the distributions diverges with $n$ for some value. Since $\Pjoint(\x,\x')$ is equal to $Q^n(\x)Q^n(\x')$ times another term, the derivative is infinite only if such term goes to $0$ fast enough so that the double sum grows asymptotically faster than $n$.

We now study the behaviour of $E_{\rm x}^n(\lambda,Q^n)$ for $\lambda\rightarrow\infty$. Since  for all $n$ the derivative is non-negative, $E_{\rm x}^n(\lambda,Q^n)$ can either converge to a constant or diverge. By calculating the limit we obtain
\begin{align}\label{eqn:592_c_gen}
\lim_{\lambda\rightarrow\infty} E_{\rm x}^n(\lambda,Q^n) &= -\frac{1}{n}\sum_{\x}\sum_{\x'}Q^n(\x)Q^n(\x') \log Z_n(\x,\x').
\end{align}
Note that the right-hand side of \eqref{eqn:592_c_gen} can be $\infty$ in some case depending on the distribution of  $\log Z_n(\X,\X')$.

Now let us consider expression \eqref{eqn:expu} and study the cases in which, for $R>0$, the optimal $\hat\lambda_n$ grows unbounded as $n\rightarrow \infty$. This happens for all rates $R<R_{\infty}(Q^n)$, where $R_{\infty}(Q^n)>0$ is a quantity that can be found following a similar reasoning as in \cite[Sec.~5.7]{gallagerBook}. Let us study the $R$-axis intercept of the following linear function 
$E_{\rm x}^n(\lambda,Q^n) + R - \lambda 2R$, given by
\begin{align}\label{eqn:R_inf_frac_gen}
R_n=\frac{E_{\rm x}^n(\lambda,Q^n)}{2\lambda - 1}.
\end{align}
Using \eqref{eqn:deriv2_1}, \eqref{eqn:deriv3_1}, \eqref{eqn:psi_def}, and taking the limit for $\lambda\rightarrow\infty$ we have
\begin{align}\label{eqn:R_inf_gen}
R_{\infty}(Q^n) &= \lim_{n \rightarrow\infty} R_n \\ \label{eqn:R_inf2}
&=\lim_{n \rightarrow\infty} -\frac{1}{2}\frac{1}{n}\log\sum_{\x}\sum_{\x'}Q^n(\x)Q^n(\x')\mathds{1}\{Z_n(\x,\x')>0\}\\ \label{eqn:R_inf3_gen}
&=\lim_{n \rightarrow\infty} -\frac{1}{2}\frac{1}{n}\log\mathbb{P}\left[Z_n(\x,\x')>0\right].
\end{align}
Expression \eqref{eqn:R_inf3_gen} is (one half) the exponent of the probability that the Bhattacharyya bound on the pairwise error probability be positive.  For all rates smaller than $R_{\infty}(Q^n)$ as given in \eqref{eqn:R_inf3_gen}, the quantity $E_{\rm x}^n(\lambda_n,Q^n) - \lambda_n 2R +R$ is infinite, which implies that $E_{\rm x}^n(\lambda_n,Q^n)$ is given by \eqref{eqn:592_c_gen} because $\hat \lambda$ is $\infty$. Notice that this is achieved using a $\hat\lambda_n$ that tends to infinity as $n$ grows. In order to assess the behaviour of the \ac{TRC} exponent we need to study the $\delta_n$ in Definition \ref{def:1} as $\hat\lambda_n\rightarrow\infty$. We recall that $\delta_n = \frac{\hat\lambda_n}{n} \log \gamma_n$. Repeating the derivation of $R_{\infty}(Q^n)$ including $\delta_n$ in the calculation, Equation~\eqref{eqn:R_inf_frac_gen} becomes (ignoring the $-1$ at the denominator which has no impact asymptotically):
\begin{align}\label{eqn:R_inf_frac_1_gen}
R_n=\frac{E_{\rm x}^n(\lambda,Q^n) -\lambda\frac{\log \gamma}{n} }{2\lambda} =
\frac{E_{\rm x}^n(\lambda,Q^n)}{2\lambda} - \frac{\log \gamma}{2n} .
\end{align}
Taking the limit in \eqref{eqn:R_inf_frac_1_gen} we obtain again $R_\infty(Q^n)$ as in~\eqref{eqn:R_inf3_gen}. If $R_{\infty}(Q^n)>0$, for all rates $R<R_{\infty}(Q^n)$ the \ac{TRC} exponent is infinite. Note that in such case  it can happen that $\delta_n \to  \infty$, as $\lambda_n$ must grow asymptotically like $E_{\rm x}^n(\lambda,Q^n)$, which could be actuall faster than $\frac{n}{\log\gamma_n}$. However, also in such case the statement of Theorem \ref{theo:2} would give a non-trivial result, as $E_{\rm x}^n(\lambda,Q^n)$ grows faster than $\delta_n$.
The $R_{\infty}(Q^n)$ derived here is closely related to the quantity $R_{\rm x}^{\infty}$ derived in \cite[Sec.~5.7]{gallagerBook} for the \ac{DMC} and is, in fact, equal to $1/2$ times the equivalent of $R_{\rm x}^{\infty}$ for general channels and pairwise-independent code ensembles. Similarly to $R_{{\rm x},\infty}$, the quantity $2R_{\infty}(Q^n)$ is a lower bound on the zero-error capacity for general channels. This implies that for all code ensembles and channels considered in the present section for which the zero-error capacity is $0$, $R_{\infty}(Q^n)=0$ as well. 

Let us now consider the case in which $R_{\infty}(Q^n)=0$.
Expression \eqref{eqn:R_inf3_gen} implies $R_{\infty}(Q^n)=0$ whenever the probability that the Bhattacharyya bound on the codeword transition probability decreases sub-exponentially. To understand what happens to the \ac{TRC} exponent in such case, we need to check both \eqref{eqn:592_c_gen}, which is the average exponent of the bound, and the behaviour of $\delta_n$. Note that expression \eqref{eqn:592_c_gen} does not necessarily go to $\infty$ when $R_{\infty}(Q^n)=0$.
For strictly zero rates, namely when $\lim_{n\to \infty} \frac 1n \log M_n = 0$, the largest exponent in~\eqref{eqn:expu} is achieved for $\hat\lambda_n\to \infty$. We then consider the dependence on $n$ in the rate by setting $R=\frac 1n \log M_n$ in~\eqref{eq:27} and let $\hat\lambda_n$ be
\begin{equation}
	\hat\lambda_n = \argmax_{\lambda\geq 1} \biggl\{ \Exnvarrho  - \frac{2\lambda}{n}\log M_n\biggr\}.
	\label{eq:272}
\end{equation}
In order to characterize how $\hat\lambda_n$ grows as $n\to\infty$, we find the Taylor series expansion of $\Exn$ around $\lambda\to\infty$, yielding
\begin{equation}
	\Exnvarrho = \nu_0^n(Q^n) - \frac{\nu_1^n(Q^n)}{\lambda} + O(\lambda^{-2}),
	\label{eq:exxx}
\end{equation}
where 
\begin{equation}
	\nu_0^n(Q^n) =  -\frac{1}{n} \mathbb{E}[\log Z_n(\X,\X')]
	\label{eq:a0}
\end{equation}
and
\begin{align}
	\nu_1^n(Q^n) = \frac{1}{2n} {\rm Var}[\log Z_n(\X,\X')].
	\label{eq:a1}
\end{align}
Then, it follows that $\hat\lambda_n$ grows as
\begin{equation}\label{eqn:rho_infty_gen}
	\hat{\lambda}_n =  \sqrt{\frac{\nu_1^n(Q^n)}{2 \frac{\log M_n}{n} +\frac{\log\gamn}{n}}} + \sigma_n,
\end{equation}
where the two terms at the denominator vanish with $n$, and also $\sigma_n \to  0$. From \eqref{eqn:rho_infty_gen} it follows that $\delta_n$ defined in~\eqref{eq:delta} satisfies $\lim_{n\to\infty}\delta_n = 0$ for $R=0$ if $\frac 1n {{\rm Var}[Z_n(\X,\X')]}$ grows slower than $\sqrt{\tfrac{n}{\log\gamma_n}}$. Note that ${{\rm Var}[\log Z_n(\X,\X')]}$ depends both on the channel and the specific code ensemble. The parameter $\nu_0^n(Q^n)$ in \eqref{eq:a0} is the same as the expression in \eqref{eqn:592_c_gen}. Note that the analysis on the growth rate of $\hat{\lambda}_n$ just carried out applies to all cases in which $\hat{\lambda}_n\rightarrow\infty$. Thus, apart from the case $R=0$, it also holds for $0<R<R_{\infty}(Q^n)$. In such case the denominator of \eqref{eqn:rho_infty_gen} does not vanish with $n$. This implies that ${\rm Var}[\log Z_n(\X,\X')] \to \infty$ faster than $n$. The term $\nu_1^n$ in~\eqref{eq:a1} characterizes the backoff of the error exponent~\eqref{eq:exxx} from its limiting value as $\lambda\to\infty$.

%%%%%%%%%%%%%%%%%%%%%%%%%%%%%%%%%%%
\section{Memoryless Channels}\label{sec:spec_ensembles}
In this section, we specialize Theorem~\ref{theo:2} to  memoryless channels with i.i.d., constant-composition and cost-constrained ensembles. As a by-product, we provide a direct dual-domain derivation of the expurgated exponent for constant-composition-codes \cite{czisar_TIT1981}, as an alternative to that  in \cite{sccarlett_TIT2014}.

%%%%%%%%%%%%%%%%%%%%%%%%%%%%%%%%%%%

\subsection{i.i.d. Ensemble}
Let us consider the i.i.d.~ensemble with distribution
\beq
Q^n _{\rm iid}(\x) = \prod_{i=1}^n Q(x_i).
\label{eq:iid}
\eeq
Using~\eqref{eq:iid} in~\eqref{eqn:expu}, we recover the single-letter version of the Gallager's $E_{\rm x}$-function~\cite[Eq.~(5.7.12)]{gallagerBook}, namely
\begin{equation}
\Exiidn = -\lambda\log\sum_{x} \sum_{x'} Q(x)Q(x') Z(x,x')^{\frac{1}{\lambda}},
\end{equation}
where $Z(x,x')$ is the single-letter  Bhattacharyya coefficient, and the i.i.d.~expurgated exponent in~\eqref{eqn:ex} becomes
\begin{equation}
	\Eexiid =   E_{\rm x}^{\rm iid}(\hat \lambda, Q) - \hat\lambda R.
	\label{eq:26}
\end{equation}
When $R>0$, it follows that the optimal parameter~\eqref{eq:rhohatn} does not depend on $n$, its limit exists and reads 
\begin{equation}
	\hat\lambda = \argmax_{\lambda\geq 1} \bigl\{ E_{\rm x}^{\rm iid}(\lambda, Q)  - \lambda 2R\bigr\}.
	\label{eq:27}
\end{equation}
In all cases for which $\hat\lambda < \infty$, the condition $\lim_{n\to\infty} \delta_n = 0$ of the theorem is satisfied. For all cases in which $\hat\rho$ is infinite we can specialize the discussion in Section \ref{sec:unbounded_rho} to the i.i.d. case. For strictly zero rates, namely when $\lim_{n\to \infty} \frac 1n \log M_n = 0$, the largest exponent in~\eqref{eq:26} is achieved for $\hat\lambda\to \infty$. 
 Then, from \eqref{eqn:rho_infty_gen}, it follows that $\hat\lambda_n$ grows as
\begin{equation}\label{eqn:rho_infty}
	\hat{\lambda}_n =  \sqrt{\frac{n\cdot \nu_1(Q)}{2 \log M_n+\log\gamn}} + \sigma_n,
\end{equation}
where
\begin{align}
	\nu_1(Q) = \frac 12{\rm Var} \left[\log  Z(x,x')\right]
	\label{eq:a1}
\end{align}
is constant in $n$ while $\sigma_n$ is a term that vanishes as $n\to\infty$, implying that $\delta_n$ defined in~\eqref{eq:delta} satisfies $\lim_{n\to\infty}\delta_n = 0$ as in the $R>0$ case.

The case in which $\hat{\lambda}_n$ grows unbounded for positive rates is obtained by replacing $R_{\infty}(Q^n)$ (after optimization over $Q^n$) with $\frac{1}{2}R_{\rm x}^{\infty}$ \cite{gallagerBook} in Section~\ref{sec:unbounded_rho}. As mentioned for the general case, also for the \ac{DMC} the $\delta_n$ can, in principle, diverge when $R<R_{\infty}(Q^n)$, and yet $\Etrclbmax\rightarrow\infty$.

According to Theorem~\ref{theo5}, with probability approaching one as $n\to\infty$, an i.i.d.~code $\Cns^{\rm iid}$ randomly generated from the ensemble~\eqref{eq:iid} has an error exponent~\eqref{eq:error1} satisfying
\begin{equation}
	E_n(\Cns^{\rm iid}) \geq E_{\rm ex}^{\rm iid}(2R,Q) + R - \delta_n
	\label{eq:32}
\end{equation}
with vanishing $\delta_n$.  Particularizing~\eqref{eq:26} to the \ac{BSC} case, it follows that the bound~\eqref{eq:32} coincides with the TRC exponent given by Barg and Forney in \cite{barg_forney_TIT2002}. 

\begin{remark}
Let us denote the smallest rate for which $\hat{\lambda}=1$ with $R^*$  and recall that the expurgated exponent and the random-coding exponent coincide for such rate \cite[pg. 154]{gallagerBook}. Similarly as in Remark \ref{rem:4}, it can be shown that, for all cases in which $\delta_n \to  0$, Theorem \ref{theo:2} implies that the achievable typical random-coding exponent is lower bounded as follows:
\begin{align}
\Etrc \geq 
\begin{cases}
E_{\rm ex}(2R,Q)+R &  0\leq R\leq \frac{R*}{2}\\
\Er &  R>\frac{R*}{2}.
\end{cases}
\end{align}
Note that $\Er$ might become zero above a certain rate. If a maximization with respect to the input distribution is carried out, such rate coincides with the capacity of the channel.
This lower bound is known to be tight for constant-composition ensembles~\cite{merhav_TIT2018} and coincides with the one derived in \cite{barg_forney_TIT2002} for the \ac{BSC}. 
\end{remark}

%%%%%%%%%%%%%%%%%%%%%%%%%%%%%%%
\subsection{Constant-Composition Ensemble}
For every $n$, let $\hat Q_n$ be a type, or empirical distribution, such that $\|\hat Q_n-Q\|_\infty\leq \frac 1n$ where $\|P\|_\infty = \max_x P(x)$. Then, the constant-composition ensemble has codeword distribution
\beq
Q^n _{\rm cc}(\x) = \frac{1}{|\Tc^n(\hat Q_n)|} \mathds{1}\big\{\x\in\Tc^n(\hat Q_n) \big\},
\label{eq:distcc}
\eeq
where $\Tc_n(\hat Q_n)$ is the type class, i.e., the set of all sequences of length $n$ with empirical distribution $\hat Q_n$, while $\mathds{1}\{.\}$ is the indicator function.

For the constant-composition ensemble~\eqref{eq:distcc}, the normalized multi-letter Gallager's expurgated function $\Exn$ reads
\begin{align}
\Exn = -\frac{1}{n} \! \log \biggl(\sum_{\x}\! Q^n(\x)\! \sum_{\x'}\! Q^n(\x')\! \prod_{i=1}^nZ(x_i,x_i')^{\frac{1}{\lambda}}\! \biggr)^\lambda.
\label{eq:exncc}
\end{align}
In~\eqref{eq:exncc}, we have used the fact that the channel is memoryless to express $Z_n(\x,\x')$ as a product, and we have also changed the order of the summations over $\x$ and $\x'$.

Since all codewords $\x'$ have the same probability $\frac{1}{|\Tc^n(\hat Q_n)|}$, the summation over $\x'$ in \eqref{eq:exncc} for a fixed constant-composition sequence $\x$ satisfies
\begin{align}
\sum_{\x'}Q^n(\x')\prod_{i=1}^nZ(x_i,x_i')^{\frac{1}{\lambda}}&=\frac{1}{|\Tc^n(\hat Q_n)|}\sum_{\x'}\prod_{i=1}^nZ(x_i,x_i')^{\frac{1}{\lambda}}\label{eqn:polty_1}\\
&\leq \min_{\bar P} \biggl\{(n+1)^{|\Xalpha|-1}e^{nD(\hat Q_n\|\bar P)}\prod_{i=1}^n\sum_{{x}'}\bar P(x')Z(x_i,x')^{\frac{1}{\lambda}}\biggr\},
\label{eqn:polty_2}
\end{align}
where the upper bound in \eqref{eqn:polty_2} follows by identifying $g_i(x_i)=Z(x_i,x_i')^{\frac{1}{\lambda}}$ in \cite[Eq.~(2.4)]{poltyrev_1982}, $\bar P$ is an auxiliary probability distribution and $D(Q \|\bar P)$ is the relative entropy between distributions $Q$ and $\bar P$. Using the upper bound~\eqref{eqn:polty_2} into~\eqref{eq:exncc} and arranging terms, we obtain
\begin{align}
\Exn& \geq  \max_{\bar P} \biggl\{ -\frac 1n \log \biggl( (n+1)^{|\Xalpha|-1}e^{nD(\hat Q_n\|\bar P)}\frac{1}{|{\cal T}^n(\hat Q_n)|} \sum_{\x} \prod_{i=1}^n\sum_{{x}'}\bar P(x')Z(x_i,x')^{\frac{1}{\lambda}}\biggr)^\lambda  \biggr\}
\label{eq:exncc3}\\
&=\max_{\bar P} \biggl\{ -\frac 1n \log \biggl(  (n+1)^{|\Xalpha|-1}e^{nD(\hat Q_n\|\bar P)} \prod_{x}\biggl(\sum_{{x}'}\bar P(x')Z(x,x')^{\frac{1}{\lambda}} \biggr)^{n\hat Q_n(x)} \biggr)^\lambda \biggr\},
\label{eq:exncc4}
\end{align}
where \eqref{eq:exncc4} follows since the sum over $\x$ in~\eqref{eq:exncc3} has exactly $|\Tc^n(\hat Q_n)|$ terms and from elementary properties of constant-composition sequences. Next, we rewrite the maximization over $\bar P$ as maximization over an auxiliary function $a(x)$  \cite{sccarlett_TIT2014} satisfying
$\bar P(x)= \hat Q_n(x)e^{ \frac{a(x)}{\lambda}}$. 
Using this in~\eqref{eq:exncc4} we find that
\begin{align}
\Exn &\geq  \max_{a(\cdot)} \biggl\{ -\frac 1n \log \biggl(  (n+1)^{|\Xalpha|-1}e^{-\frac{n}{\lambda} \sum_x\hat Q_n(x) a(x)}  \prod_{x}\biggl(\sum_{{x}'} \hat{Q}_n(x')e^{ \frac{a(x')}{\lambda}} Z(x,x')^{\frac{1}{\lambda}} \biggr)^{n\hat Q_n(x)}\biggr)^\lambda \biggr\}
\label{eq:exncc5}\\
&=- \frac{\lambda(|{\cal X}|+1)}{n} \log(n+1) +  \max_{a(\cdot)} \biggl\{ - \lambda \log  \prod_{x}\biggl(\sum_{{x}'} \hat Q_n(x') \Bigl(  Z(x,x') e^{ a(x')-a(x)}\Bigr)^{\!\frac{1}{\lambda}} \biggr)^{\!\!\hat Q_n(x)} \biggr\}.
\label{eq:exncc6}
\end{align}

We observe that the the first term in the r.h.s.~of~\eqref{eq:exncc6} vanishes with $n$. Using that $\|\hat Q_n-Q\|_\infty\leq \frac 1n$, and further simplifying the result, we obtain the constant-composition version of the Gallager's $E_{\rm x}$-function, namely
\begin{align}
E_{\rm x}^{\rm cc}(\lambda,Q) = &\max_{a(\cdot)}\left\{-\lambda \sum_x Q(x)\log \sum_{{x}'}Q(x')\Bigl(Z(x,x')e^{ a(x')-a(x)}\Bigr)^{\frac{1}{\lambda}}\right\}.
\end{align}
According to Theorem~\ref{theo5}, with probability approaching one as $n\to\infty$, a constant-composition code $\Cns^{\rm cc}$ randomly generated from the ensemble~\eqref{eq:distcc} has an error exponent satisfying
\begin{equation}
	E_n({\cal C}_n^{\rm cc}) \geq \Eexxcc +R,
\end{equation}
where 
\beq
\Eexcc = E_{\rm x}^{\rm cc}(\hat\lambda,Q) -\hat\lambda R
\eeq
is the constant-composition version of the expurgated exponent with optimal parameter
\begin{equation}
	\hat\lambda = \argmax_{\lambda\geq 1} \bigl\{ E_{\rm x}^{\rm cc}(\lambda, Q)  - \lambda 2R\bigr\},
	\label{eq:27x}
\end{equation}
coinciding with the expression in \cite{merhav_TIT2018}. Similarly to what seen for the i.i.d. case, we need to address the case in which $\hat\lambda_n$ diverges as $n$ grows. For the zero-rate case we find that the optimal $\hat\lambda_n$ grows as
\begin{equation}\label{eqn:rho_infty}
	\hat{\lambda}_n =  \sqrt{\frac{n\cdot \nu_1(Q)}{2 \log M_n+\log\gamn}} + \sigma_n,
\end{equation}
where
\begin{align}
	\nu_1(Q) = \max_{a(\cdot)} \frac 12{\rm Var} \left[\log \left(Z(x,x')e^{ a(x')-a(x)}\right)\right],
	\label{eq:a1}
\end{align}
impliying that $\delta_n\rightarrow 0$.

The case in which $\hat{\lambda}_n$ grows unbounded for positive rates is obtained in a similar way as for the i.i.d. case by using the equivalent of $R_{\rm x}^{\infty}$ for the constant-composition case, which can be derived in a similar way as done for the general case in Section \ref{sec:unbounded_rho} and the i.i.d. case in \cite[Sec.~5.7]{gallagerBook}.

At high rates the typical error exponent coincides with the random-coding exponent for constant-composition ensembles \cite{scarlettAllerton_ens_tight2012} %\footnote{The random coding exponent for constant-composition codes can be also derived starting from the results in \cite{poltyrev_1982}.}:
\begin{align}
E_{\textrm{r}}^{\textrm{cc}}(Q,R)=\max_{\overline{Q}} E_0(\overline{Q},\rho) -(1-\rho)D(Q\|\overline{Q}) - \rho R,
\end{align}
where $E_0(\overline{Q},\rho)$ is Gallager's $E_0$ function for the \ac{i.i.d.} ensemble while $D(Q\|\overline{Q})$ is the relative entropy between the distributions $Q$ and $\overline{Q}$.

%%%%%%%%%%%%%%%%%%%%%%%%%%%%%
\subsection{Cost-Constrained Ensemble}
The cost-constrained ensemble is described by a set of symbol-wise cost functions and an i.i.d. input distribution $Q^n(\x)=\prod_{i=1}^n Q(x_i)$ that induce the pairwise-independent codeword distribution:
\begin{equation}
	Q^n_{\rm cost}(\x) = \frac{1}{\mu_n} \prod_{i=1}^n Q(x_i)\mathds{1}\left\{ \x \in {\cal D}_n\right\},
	\label{eq:Qcost}
\end{equation}
where $\mu_n$ is a normalization factor and ${\cal D}_n$ is the set of input sequences satisfying the cost constraints given by
\begin{equation}\label{eqn:cost_set_def}
	{\cal D}_n = \left\{  \x \, : \,  \left| \frac{1}{n} \sum_{i=1}^n a_{\ell}(x_i) - \phi_\ell \right| \leq \frac{\delta}{n} , \, \ell=1,\ldots,L\right\}
\end{equation}
for some $\delta>0$ and $\phi_\ell= \sum_{x}Q(x) a_{\ell}(x)$  where $\{a_{\ell}(x)\}_{\ell=1}^L$ are auxiliary cost functions and can be optimized to obtain an improved exponent with respect to the i.i.d. case. A system cost of the form $\frac{1}{n} \sum_{i=1}^n c(x_i) \leq \Gamma $ can be included in the analysis, except that no optimization over the function $c(x)$ itself can be done. This ensemble \cite{sccarlett_TIT2014} is a generalization of the one in \cite[Sec.~7.3]{gallagerBook}, which considers a single constraint.
As done for the \ac{i.i.d.} and the constant-composition ensembles, we specialize the result of Theorem \ref{theo:2} to the cost constrained case by upper bounding $\Etrclbl$ and $\Etrclbh$. The key advantage of this ensemble is that it allows to obtain the same performance of constant-composition codes for channels with possibly continuous alphabets.

Let us start with $\Etrclbl$.
Plugging~\eqref{eq:Qcost} into~\eqref{eqn:ex} we obtain
\begin{equation}\label{eq:Ecost}
E_{\rm x}^n(\lambda,Q^n_{\rm cost}) = -\frac 1n \log \left(\frac{1}{\mu_n^2} \sum_{\x} \sum_{\x'} Q^n(\x) Q^n(\x') \mathds{1}\left\{ \x \in {\cal D}_n \right\} \mathds{1}\left\{ \x' \in {\cal D}_n \right\} Z_n(\x,\x')^\frac{1}{\lambda} \right)^\lambda.
\end{equation}
Under mild assumptions on the cost functions, namely 
%$\mathbb{E}_Q\left[c(X)\right] \leq \Gamma$, $\mathbb{E}_Q\left[c(X)^2\right] \leq \infty$ 
and $\mathbb{E}_Q\left[a_\ell(X)^2\right] \leq \infty$ for all $\ell=1,2,\ldots,L$, the normalization constant satisfies~\cite[Prop.~2]{sccarlett_TIT2014},\cite{stone1965_local_ratio_theo} $\lim_{n\to\infty}\frac 1n\mu_n = 0$, implying that, for $n$ large,
\begin{equation}
E_{\rm x}^n(\lambda,Q^n_{\rm cost}) = -\frac 1n \log \left( \sum_{\x} \sum_{\x'} Q^n(\x) Q^n(\x') \mathds{1}\left\{ \x \in {\cal D}_n \right\} \mathds{1}\left\{ \x' \in {\cal D}_n \right\} Z_n(\x,\x')^\frac{1}{\lambda} \right)^\lambda.
\label{eq:cost2}
\end{equation}
We next find an upper bound on the indicator function in~\eqref{eq:cost2}. Define $a^n_{\ell}(\x)= \sum_{i=1}^n a_{\ell}(x_i)$. % and $c^n(\x)=\sum_{i=1}^n x(c_i)$. 
For any real-valued number $r_\ell$,  we have
\begin{align}
	\mathds{1}\left\{ \x \in {\cal D}_n \right\} &= \prod_{\ell=1}^L \mathds{1}\bigl\{ |a^n_{\ell}(\x)-n\phi_\ell | \leq \delta \bigr\}	\\
		 & \leq \prod_{\ell=1}^L e^{r_\ell(a^n_{\ell}(\x)-n\phi_\ell)} e^{|r_\ell|\delta} \label{eq:step1}\\
	 &= e^{\sum_{\ell=1}^L  r_\ell(a^n_{\ell}(\x)-n\phi_\ell) + |r_\ell|\delta},
	 \label{eq:121}
\end{align}
where in~\eqref{eq:step1} we upper bounded each indicator function using $\eqref{eqn:cost_set_def}$ for any real number $r_\ell$ \cite[Eq.~(56)]{sccarlett_TIT2014}, which is independent of $n$. Using \eqref{eq:121} we bound the product of indicator functions in    \eqref{eq:Ecost} as:  
\begin{align}\label{eq:bound_1_prod}
	\mathds{1}\left\{ \x \in {\cal D}_n \right\}\mathds{1}\left\{ \x' \in {\cal D}_n \right\} &\leq e^{\sum_{\ell=1}^L  r_\ell(a^n_{\ell}(\x)-n\phi_\ell) + |r_\ell|\delta}e^{\sum_{\ell=1}^L  \overline{r}_\ell(a^n_{\ell}(\x')-n\phi_\ell) + |\overline{r}_\ell|\delta}\\\label{eq:bound_2_prod}
	& = \frac{e^{\sum_{\ell=1}^L  r_\ell(a^n_{\ell}(\x)-n\phi_\ell) }}{e^{\sum_{\ell=1}^L  {r}_\ell'(a^n_{\ell}(\x')-n\phi_\ell) }}e^{\delta\sum_{\ell=1}^L(|r_\ell| + |{r}_\ell'|)}
\end{align}
where in \eqref{eq:bound_2_prod} we renamed $\overline{r}$ as $-r'$.
  Since  $r_\ell$ is a real-valued variable which does not depend on $n$, we can bound the argument of the logarithm in \eqref{eq:cost2} using~\eqref{eq:bound_2_prod}:
  \begin{align}\label{eq:bound_Ecost1}
\biggl( \sum_{\x} \sum_{\x'} Q^n(\x) Q^n(\x') & \mathds{1}\left\{ \x \in {\cal D}_n \right\} \mathds{1}\left\{ \x' \in {\cal D}_n \right\} Z_n(\x,\x')^\frac{1}{\lambda} \biggr)^\lambda \\
 &\leq\left( \sum_{\x} \sum_{\x'} Q^n(\x) Q^n(\x') \frac{e^{\sum_{\ell=1}^L  r_\ell'(a^n_{\ell}(\x')-n\phi_\ell) }}{e^{\sum_{\ell=1}^L  {r}_\ell(a^n_{\ell}(\x)-n\phi_\ell) }}e^{\delta\sum_{\ell=1}^L(|r_\ell'| + |{r}_\ell|)} Z_n(\x,\x')^\frac{1}{\lambda} \right)^\lambda
 \\\label{eq:bound_Ecost2}
 &\dot{=}
 \left( \sum_{\x} \sum_{\x'} Q^n(\x) Q^n(\x') \frac{e^{\sum_{\ell=1}^L  r_\ell'(a^n_{\ell}(\x')-n\phi_\ell) }}{e^{\sum_{\ell=1}^L  {r}_\ell(a^n_{\ell}(\x)-n\phi_\ell) }} Z_n(\x,\x')^\frac{1}{\lambda} \right)^\lambda,
\end{align}
where in \eqref{eq:bound_Ecost2} we used the fact that $e^{\delta\sum_{\ell=1}^L(|r_\ell| + |{r}_\ell'|)}$ is independent of $n$  while $Q^n(\x)=\prod_{i=1}^n Q(x_i)$  is an input distribution that satisfies the  conditions of \cite[Prop.~2]{sccarlett_TIT2014} as well as the system cost constraint%\footnote{The expression can be rewritten by including the system cost $c(.)$ in a similar way.}
. From \eqref{eq:bound_Ecost2} we can bound $E_{\rm x}^n(\lambda,Q^n_{\rm cost})$ as follows:
\begin{align}
E_{\rm x}^n(\lambda,Q^n_{\rm cost})\geq E_{\rm x}^n(\lambda,Q^n,\{a_\ell\},\{r_\ell\},\{r_\ell'\})=-\frac{1}{n}\log \left( \sum_{\x} \sum_{\x'} Q^n(\x) Q^n(\x') \frac{e^{\sum_{\ell=1}^L  r_\ell'(a^n_{\ell}(\x')-n\phi_\ell) }}{e^{\sum_{\ell=1}^L  {r}_\ell(a^n_{\ell}(\x)-n\phi_\ell) }} Z_n(\x,\x')^\frac{1}{\lambda} \right)^\lambda.
\end{align} 
Expanding each term as single-letter product and optimizing over the auxiliary variables, we obtain:
\begin{equation}
E_{\rm x}^n(\lambda,Q^n_{\rm cost}) \geq E_{\rm x}(\lambda,Q) =  \sup_{\{r_\ell\}_{\ell=1}^L,\{r_\ell'\}_{\ell=1}^L, \{a_{\ell}(.)\}_{\ell=1}^L} - \lambda \log \left( \sum_{x} \sum_{x'} Q(x) Q(x') \frac{e^{\sum_{\ell=1}^L r_\ell'(a_{\ell}(x')-\phi_\ell)}}{e^{\sum_{\ell=1}^L r_\ell(a_{\ell}(x)-\phi_\ell)}} Z(x,x')^\frac{1}{\lambda} \right).
\label{eq:123}
\end{equation}

From \eqref{eq:123}, the bound $\Etrclbl$ follows. As a side remark we point out  that the expurgated of the \ac{i.i.d.} and the constant-composition ensembles can be recovered from \eqref{eq:123}. The former is recovered by setting $r_\ell=r_\ell'=0$ in \eqref{eq:123}, while the latter is obtained by setting $L=1$, $r_1'=1$ and $r_1=0$. This differs from the similar result in \cite{sccarlett_TIT2014}, where the constant-composition exponent is obtained with $L=2$, due to the fact that \cite{sccarlett_TIT2014} considers a mismatched decoder rather than a \ac{ML} one as in our case.

Let us now consider the case in which $\hat\lambda_n\to\infty$  as $n$ grows. For the zero-rate case we find from \eqref{eqn:rho_infty_gen} that
\begin{equation}\label{eqn:rho_infty_cost}
	\hat{\lambda}_n =  \sqrt{\frac{n\cdot \nu_1(Q)}{2 \log M_n+\log\gamn}} + \sigma_n,
\end{equation}
where now the term $\nu_1(Q)$ is given by
\begin{align}
	\nu_1(Q) = \sup_{\{r_\ell\}_{\ell=1}^L,\{r_\ell'\}_{\ell=1}^L, \{a_{\ell}(\cdot)\}_{\ell=1}^L} \frac 12{\rm Var} \left[\log \left(\frac{e^{\sum_{\ell=1}^L r_\ell'(a_{\ell}(x')-\phi_\ell)}}{e^{\sum_{\ell=1}^L r_\ell(a_{\ell}(x)-\phi_\ell)}} Z(x,x')\right)\right],
	\label{eq:a1_cost}
\end{align}
implying that $\delta_n\rightarrow 0$. 

The case in which $\hat{\lambda}_n$ grows unbounded for positive rates is obtained in a similar way as for the \ac{i.i.d.} and the constant-composition ensembles by using the equivalent of $R_{\rm x}^{\infty}$ for the cost constraint case.

As for the bound $\Etrclbh$, it can be obtained by plugging~\eqref{eq:Qcost} into~\eqref{eqn:rce}, using the bound on the indicator function given in \eqref{eq:121} and following  a similar reasoning as for $E_{\rm x}^n(\lambda,Q^n_{\rm cost})$ to obtain \cite{sccarlett_TIT2014,scarlettAllerton_ens_tight2012}
 \begin{align}\label{eq:129}
E_0^n(\rho,Q^n_{\text{cost}})\geq E_0^n(\rho,Q)= \sup_{\{r_\ell\}_{\ell=1}^L,\{r_\ell'\}_{\ell=1}^L, \{a_{\ell}(.)\}_{\ell=1}^L}\log\Bigg(\sum_{y}\bigg(\sum_{x}Q(x)e^{\sum_{\ell=1}^L  r_\ell(a_{\ell}(x)-\phi_\ell)}\Wyx^{\frac{1}{1+\rho}}\bigg)^{1+\rho}\Bigg),
\end{align}
from which the bound on the \ac{TRC} exponent at high rates $\Etrclbh$ follows.

%%%%%%%%%%%%%%%%%%%%%%%%%%%%%%%%
\section{Finite-State Channels}\label{sec:memory}

In this section, we particularize the result in Theorem \ref{theo:2} to the case of finite-state channels. In particular, we derive two lower bounds on the exponents $\Etrclbl$ and $\Etrclbh$ in~\eqref{eqn:theo5_statement2}. 

 We consider finite-state channels with $A$ states $\{1,2,\ldots,A\}$. The statistical behaviour of the channel is described by a joint probability measure linking the current channel state and output to the current input and previous state $p(y_t,s_t|x_t,s_{t-1})$. Similarly to~\cite[Sec.~4.6]{gallagerBook}, our results are based on the probability of the channel output $\y$, given the channel input $\x$ and initial state $s_0$, which can be obtained by summing all the possible state sequences $(s_1,\ldots,s_n)$ using the following recursive expression,
\begin{align}
W^n(\y|\x,s_0) =\sum_{s_n=1}^A p_n(\y,s_n|\x,s_0),
\label{eqn:memo_init_state}
\end{align}
where the probability $p_n(\y,s_n|\x,s_0)$ can be calculated using the recursion
\begin{align}
p_t(\y_t,s_t|\x_t,s_0)= \sum_{s_{t-1}=1}^A p(y_t,s_t|x_t,s_{t-1})  p_{n-1}(\y_{t-1},s_{t-1}|\x_{t-1},s_0),
\label{eqn:state}
\end{align}
for $t=1,\ldots,n$. In~\eqref{eqn:state}, $\x_t$ and $\y_t$ are respectively the channel input and output sub-sequences given by  $\x_t=(x_1,\ldots,x_t)$ and $\y_t=(y_1,\ldots,y_t)$, while for $t=1$ we have the initial  relation $p_1(y_1,s_1|x_1,s_0) = p(y_1,s_1|x_1,s_0)$. Equation~\eqref{eqn:memo_init_state} represents a family of channel transition probabilities, indexed by the initial state $s_0$, that also encompasses compound channels since $s_0$ can be considered as an index that determines the channel from the compound set.
We study the error probability of ensemble of codes using a mismatched decoder that employs a decoding metric that  averages over the initial state assuming an equiprobable state distribution, as in \cite{gallagerBook},
\begin{align}
W^n(\y|\x) &=\sum_{s_0=1}^A \frac{1}{A}W^n(\y|\x,s_0)\label{eqn:memo_ineq_1}.
\end{align}

\subsection{Lower Bound on $\Etrclbl$ for Finite-State Channels}

From Lemma \ref{lem:main} and the error probability bound in~\eqref{eqn:trc_from_gallager_last_+}, we have that with high probability, a code drawn randomly from a pairwise-independent ensemble used over a FSC and averaged over all possible states has an error probability bounded by
\begin{align}
P_{\rm e}(\mathcal{C}_n) &\leq \frac{1}{M_n}\left(\gamma_n M_n(M_n-1)\right)^{\lambda}\Bigg(\sum_{\x}\sum_{\x'}Q^n(\x)Q^n(\x') \bigg(\sum_{\y}\sqrt{W^n(\y|\x)W^n(\y|\x')}\bigg)^\frac{1}{\lambda}  \Bigg) ^{\lambda}\\
&\leq \frac{1}{M_n}\left(\gamma_n M_n(M_n-1)\right)^{\lambda}\Bigg(\sum_{\x}\sum_{\x'}Q^n(\x)Q^n(\x') \Bigg(\sum_{\y}\sqrt{\sum_{s_0=1}^A \frac{1}{A} W^n(\y|\x,s_0)\sum_{s_0'=1}^A \frac{1}{A}W^n(\y|\x',s_0')}\Bigg)^\frac{1}{\lambda}  \Bigg) ^{\lambda},
\label{eqn:trc_mem_aver}
\end{align}
where \eqref{eqn:trc_mem_aver} follows from \eqref{eqn:memo_ineq_1}. From \eqref{eqn:trc_mem_aver} it is possible to find a bound that holds for any given initial state, useful in case a distribution over such state is unknown. Referring to the error probability for code $\mathcal{C}_n$ given an initial state $\bar{s}_0$ as $P_{\rm e}(\mathcal{C}_n, \bar{s}_0)$, we have the bounds
\begin{align}
P_{\rm e}(\mathcal{C}_n)& =\sum_{\bar{s}_0}q(\bar{s}_0)P_{\rm e}(\mathcal{C}_n, \bar{s}_0) \\ & \geq \max_{\bar{s}_0}q(\bar{s}_0) P_{\rm e}(\mathcal{C}_n, \bar{s}_0)\label{eqn:trc_mem_s0}\\
&\geq \frac{1}{A} P_{\rm e}(\mathcal{C}_n, \bar{s}_0)\label{eqn:trc_mem_s0_1}.
\end{align}
From \eqref{eqn:trc_mem_s0_1} we obtain
\begin{align}
P_{\rm e}(\mathcal{C}_n, \bar{s}_0)\leq A P_{\rm e}(\mathcal{C}_n) \label{eqn:trc_mem_s0_1_1},
\end{align}
a bound that holds independently of the initial state distribution or on whether such distribution exists or not. 
Plugging \eqref{eqn:trc_mem_aver} into \eqref{eqn:trc_mem_s0_1_1} we get an upper bound on the average error probability for a given initial state which does not depend on the initial state distribution. From Theorem \ref{theo5}, keeping the dependency on the initial state into account, we have that, with high probability, for any initial state $s_0$: 
\begin{align}
P_{\rm e}(\mathcal{C}_n, s_0) &\leq 
\frac{A}{M_n}\left(\gamma_n M_n(M_n-1)\right)^{\lambda}\Bigg(\sum_{\x}\sum_{\x'}Q^n(\x)Q^n(\x') \Bigg(\sum_{\y}\sqrt{\sum_{s_0=1}^A \frac{1}{A} W^n(\y|\x,s_0)\sum_{s_0'=1}^A \frac{1}{A} W^n(\y|\x',s_0')}\Bigg)^\frac{1}{\lambda}  \Bigg) ^{\lambda}\\
&= 
\frac{1}{M_n}\left(\gamma_n M_n(M_n-1)\right)^{\lambda}\Bigg(\sum_{\x}\sum_{\x'}Q^n(\x)Q^n(\x') \Bigg(\sum_{\y}\sqrt{\sum_{s_0=1}^A  W^n(\y|\x,s_0)\sum_{s_0'=1}^A  W^n(\y|\x',s_0')}\Bigg)^\frac{1}{\lambda}  \Bigg) ^{\lambda}
\label{eqn:trc_mem_final}
\end{align}
where in \eqref{eqn:trc_mem_final} we simplified the outer $A$ with those under the square root.
Taking the negative normalized logarithm in \eqref{eqn:trc_mem_final} we finally get a lower bound on the exponent of a typical code from an ensemble over a finite state channel with initial state $s_0$ for a given $n$. Thus \eqref{def:1} can be rewritten as: 
\begin{align}\label{def:1s0}
\Etrclbl   =  F_{\rm x}^n(\hat\lambda_n,Q^n) -  2\hat\lambda_n R - \delta_n 
\end{align}
where
\begin{align}\label{eqn:ex_s0}
F_{\rm x}^n(\lambda,Q^n) & =  -\frac{1}{n}\log \Bigg(\sum_{\x}\sum_{\x'}Q^n(\x)Q^n(\x') \Bigg(\sum_{\y}\sqrt{\sum_{s_0=1}^A  W^n(\y|\x,s_0)\sum_{s_0'=1}^A  W^n(\y|\x',s_0')}\Bigg)^\frac{1}{\lambda}  \Bigg) ^{\lambda}.
\end{align}
We next show that the limit of \eqref{eqn:ex_s0} for $n$ that tends to infinity exists and is finite. We start with the following lemma, which is the equivalent for \ac{TRC} of \cite[Lemma (5.9.1)]{gallagerBook}.
\begin{lemma}\label{lem:591}
For any finite-state channel the following holds:
\begin{align}\label{eqn:591_a}
F_{\rm x}^n(\lambda,Q^n)\geq \frac{k}{n} F_{k}(\lambda,Q^k) + \frac{l}{n} F_{l}(\lambda,Q^l) 
\end{align}
where $k$ and $l$ are positive integers and $k+l=n$.
\end{lemma}
\begin{IEEEproof}
See Appendix~\ref{ap:A}.
\end{IEEEproof}

The next lemma is the equivalent of \cite[Lemma (5.9.2)]{gallagerBook} for the typical error exponent case.
\begin{lemma}\label{lemma:3}
Let us define:
\begin{align}\label{eqn:592_a}
F_{\rm x}^{\infty}(\lambda,Q) =  \sup_n F_{\rm x}^n(\lambda,Q^n),
\end{align}
$Q$ being the limiting distribution of $Q^n$, which is assumed to exist.
For all pairwise-independent code ensembles and all finite state channels for which there exists an $n_0$ such that for $n>n_0$ the normalized relative entropy \eqref{eqn:deriv3}
\begin{align}
\frac{1}{n}D\left(\Pjoint^n(\x,\x')\|Q^n(\x)Q^n(\x')\right)<\infty
\end{align}
and for $\lambda\geq 1$ we have: 
\begin{align}\label{eqn:592_b}
\lim_{n\rightarrow\infty} F_{\rm x}^n(\lambda,Q^n) = F_{\rm x}^{\infty}(\lambda,Q).
\end{align}
Furthermore, for $1\leq \lambda <\infty$ the convergence is uniform in $\lambda$ and $F_{\rm x}^{\infty}(\lambda,Q)$ is uniformly continuous in $\lambda$.
\end{lemma}
\begin{IEEEproof}
Consider first the case $1\leq \lambda <\infty$. The boundedness and positiveness of the derivative \eqref{eqn:deriv3}, particularized to finite-state channels, for any finite $\lambda$ together with the fact that $F_{\rm x}^n(1,Q^n)$ is finite (see proof of Lemma \ref{lemma:4} below and \cite[Lemma (5.9.2)]{gallagerBook}) implies that $F_{\rm x}^n(\lambda,Q^n)$ is positive and bounded. This fact together with Lemma \ref{lem:591} allows us to apply \cite[Lemma (4A.2)]{gallagerBook}, which implies \eqref{eqn:592_b}.
Furthermore, the finiteness of the derivative in $\lambda$ for each $n$ implies uniform convergence and uniform continuity.

For the case in which $\hat\lambda_n\rightarrow\infty$, we have that the following limit:
\begin{align}\label{eqn:592_d}
\lim_{n\rightarrow\infty} F_{\rm x}^n(\hat\lambda_n,Q^n)
\end{align}
exists and is either finite or infinite. This follows from the fact that the derivative of $F_{\rm x}^n(\lambda,Q^n)$ with respect to $\lambda$ exists for any $n$ and is positive.
\end{IEEEproof}
An important implication of Lemma \ref{lem:591} is that it guarantees that the limit in \eqref{eqn:theo5_limit} exists. This means that Theorem \ref{theo:2} gives a result which is non trivial and in fact practically relevant.

The behavior of $\Etrclbmax$ in the case $\hat\lambda_n \to \infty$ can be derived from the general case presented in Section \ref{sec:main_res} by substituting the Bhattacharyya bound on the pairwise error probability with:
\begin{align}
\sum_{\y}\sqrt{\sum_{s_0=1}^A  W^n(\y|\x,s_0)\sum_{s_0'=1}^A  W^n(\y|\x',s_0')}.
\end{align}

From the discussion above it follows that the limit \eqref{eqn:592_d} is always $\infty$ for $R>0$ (the dependency on $R$ is hidden in $\hat\lambda_n$, which is in turn a function of $R$), since $\hat\lambda_n\rightarrow\infty$ for a positive rate $R$ only if $R<R_{\infty}(Q^n)$, while it can be finite in $R=0.$

%%%%%%%%%%%%%%%%%%%%%%%%%%%%%
\subsection{Lower Bound on $\Etrclbh$ for Finite-State Channels}
We recall from Definition \ref{def:2} that $\Etrclbh  =   \Ern - \iota_n$  where $\Ern$ is the random-coding exponent while, under the conditions of Theorem \ref{theo:2}, $\iota_n \to 0$.
The derivation of a lower bound to $\Ern$ for finite-state channels was developed in \cite[Sec.~5.9]{gallagerBook}, leading to the random-coding error exponent
\begin{align}\label{eqn:fsc_gallager}
E_{\rm r}(R,Q) = \max_{0\leq \rho\leq 1}  F_{0}^{\infty}(\rho,Q) - \rho R
\end{align}
where $F_0^\infty(\rho,Q)$ is defined as
\begin{align}
F_{0}^{\infty}(\rho,Q) =  \lim_{n\rightarrow \infty}F_0^n(\rho,{Q}^n)
\end{align}
and is related to the limiting $E_0$-function as
\begin{align}
%F_{0}^{\infty}(\rho,Q) = -\frac{\rho\log A}{n} + \min_{s_0} E_0^\infty(\rho,Q,s_0)
F_{0}^{\infty}(\rho,Q) =\min_{s_0} E_0^\infty(\rho,Q,s_0)
\end{align}
with
\begin{align}
E_0^\infty(\rho,Q,s_0)=\lim_{n\rightarrow\infty}-\frac{1}{n}\log \Bigg(\sum_{\y}\bigg(\sum_{\x}{Q}^n(\x)W^n(\y|\x,s_0)^{\frac{1}{1+\rho}}\bigg)^{1+\rho}\Bigg)
\end{align}

\subsection{Rate $R^*$}\label{sec:crit_rate_fin_state}
Let us now focus on indecomposable channels\footnote{See \cite[Sec.~4.6]{gallagerBook} for a definition.}. For such channels different definitions of capacity exist. In \cite[Sec.~4.6]{gallagerBook} the upper ($\overline{C}$) and lower ($\underline{C}$) capacities are considered. They differ in that the former includes a maximization over the initial state $s_0$, while the latter includes a minimization. Interestingly, for indecomposable channels the two definitions coincide, i.e., $\overline{C}=\underline{C}=C$.
In the following we show that there exists a rate $R^*$ such that below $R^*$ $\Etrclbl$ is larger than $\Etrclbh$ while the other way round is true for rates $R^*<R<\underline{C}$. % 

We start by pointing out that $E_{\rm r}(R,Q)$ is convex and strictly decreasing in $R$ for $R<\underline{C}$, as shown in \cite{gallagerBook}, and the optimal $\rho$ is a decreasing function of $R$ so that for $R$ that goes to $\underline{C}$ the optimal $\rho$ gets close to $0$ while it takes value $1$ in a continuous interval that includes the point $R=0$. As for $F_{\rm x}^n(\lambda_n,Q^n)$, we can see that $R$ and $\lambda$ play the same roles as in $E_{\rm x}(\lambda,\boldsymbol(Q))$ as defined in \cite[Sec.~5.7]{gallagerBook}, and thus $F_{\rm x}^n(\lambda_n,Q^n) -  \lambda_n R$ is convex and strictly decreasing in $R$ and the optimal $\lambda$ is also a decreasing function in $R$. Note, however, that such quantity (which is a bound on the expurgated exponent for finite-state channels) can reach $0$ at a rate which is much smaller than the capacity. Now let us call $R_{\rm cr}$ the largest rate at which the maximum of $E_{\rm r}(R,Q)$ is obtained for $\rho=1$. 

The following  lemma  is instrumental in proving Theorem \ref{theo:3}, where the \ac{TRC} exponent for all rates is presented.

\begin{lemma}\label{lemma:4}
For finite-state channels:
\begin{align}
F_0^n(1,Q^n) - \frac{\log A}{n} \leq F_{\rm x}^n(1,Q^n)\leq F_0^n(1,Q^n) + \frac{\log A}{n}.
\end{align}
Furthermore, if $F_0^n(1,Q^n)$ is finite $F_{\rm x}^n(1,Q^n)$ is finite as well.
\end{lemma}

\begin{IEEEproof}
See Appendix~\ref{ap:B}.
\end{IEEEproof}

The following theorem shows that, for finite-state channels, the limits in Theorem \ref{theo:2} exist and, if $\delta_n \to  0$, the result of the theorem is non-trivial, i.e., the typical exponent exists and is strictly positive. Furthermore, it provides the result to the minimization  in Theorem \ref{theo:2} for each rate below capacity.
\begin{theorem}\label{theo:3}
For all finite-state channels for which $\delta_n \to  0$ the following holds:
\begin{equation}\label{eqn:theo_lim_finite_sta1}
\liminf_{n\rightarrow\infty}	\min\left\{ \Etrclbl,\Etrclbh\right\} =
\begin{cases}
E_{\rm ex}(2R,Q)+R  \ \text{ for } R\leq R^* \\
E_{\rm r}(R,Q)  \ \text{ for } R > R^*
\end{cases}
\end{equation}
where $R^*=\frac{R_{\rm cr}}{2}$. Furthermore, $E_{\rm ex}(2R,Q)+R$ is strictly positive for range $R\leq R^*$ while $E_{\rm r}(R,Q)$ is strictly positive in $R^*<R<\underline{C}$ if a maximization over $Q$ is carried out.
\end{theorem}

\begin{IEEEproof}
Using Lemma \ref{lemma:4}, the fact that both $F_{\rm x}^n(\lambda,Q^n)$ and $F_0^n(\rho,Q^n)$ are increasing in $\lambda$ and $\rho$, respectively, and the fact that $\lambda\geq 1$ while $0<\rho\leq 1$, it follows for large $n$ that
\begin{equation}
\max_{\lambda\geq 1} E_{\rm x}^n(\lambda,Q^n) - \lambda R \geq \max_{0\leq \rho\leq 1}  F_0^n(\rho,Q^n) - \rho R
\end{equation} when $R\leq R_{\rm cr}$ and that
\begin{equation}
\max_{\lambda\geq 1} E_{\rm x}^n(\lambda,Q^n) - \lambda	R < \max_{0\leq \rho\leq 1}  F_0^n(\rho,Q^n) - \rho R  
\end{equation}
for $R>R_{\rm cr}$. Thus, for $R\leq R_{\rm cr}$ we obtain that
\begin{equation}
\Eexn\geq E_{\rm r}(R,Q^n)
\end{equation}
while for $R>R_{\rm cr}$
\begin{equation}
\Eexn < E_{\rm r}(R,Q^n) \ \text{ for } R > R_{\rm cr}.
\end{equation}
Finally, since  $\Etrclbl  =  \EexnRR + R - \delta_n $, for all cases such that $\delta_n \to  0$ and with reference to the statement of Theorem \ref{theo:2}:
\begin{equation}\label{eqn:theo_lim_finite_sta4}
\liminf_{n\rightarrow\infty}	\min\left\{ \Etrclbl,\Etrclbh\right\} =
\begin{cases}
\liminf_{n\rightarrow\infty}	\Etrclbl =  E_{\rm ex}(2R,Q)+R & 0\leq R\leq R^* \\
\liminf_{n\rightarrow\infty}	\Etrclbh =  E_{\rm r}(R,Q) &  R>R^*
\end{cases}
\end{equation} 
which was obtained by equating $E_{\rm ex}(2R,Q)+R$ and $E_{\rm r}(R,Q)$, from the definition of $R_{\rm cr}$ and setting $R^*=\frac{R_{\rm cr}}{2}$. At the right-hand side of \eqref{eqn:theo_lim_finite_sta4}, the quantity $E_{\rm ex}(2R,Q)+R$ is positive for rates $R\leq R^*$, while Gallager's exponent $E_{\rm r}(R,Q)$ can be made positive up to capacity by maximizing over the input distribution, which gives the theorem statement.
\end{IEEEproof}

\subsection{State Known at the Receiver}
In the following we specialize the results for the \ac{FSC} to the case in which the state of the channel is known at the receiver. We consider a channel in which the state changes at any symbol time according to a Markov chain with a finite number $A$ of states.  In particular, we consider an \ac{FSC} in which the channel state sequence $\s$ seen by a transmitted codeword is a deterministic function of the output $\y$ and the initial state $s_0$, i.e., $\s=f(\y,s_0)$. At a given  instant, the channel is modelled as a \ac{DMC}, the characteristics of which depend on the state at previous instant. This introduces memory in the channel. Channels corresponding to different states are different \ac{DMC}s, with the only constraint that the input alphabet $\mathcal{X}$ be the same. An example of such family of channels are \ac{FSC}s for which the \ac{DMC}s have the same input set but disjoint output sets, that is, the intersection of the output alphabets of any two \ac{DMC}s is empty. This allows the receiver to know what the state in previous instant was.
Knowing the channel state allows the receiver to use maximum-likelihood decoding with the transition probability $W^n(\y|\x,s_0)$ and to write~\eqref{eqn:trc_mem_final} as
\begin{align}
P_{\rm e}(\mathcal{C}_n, s_0) &\leq 
\frac{1}{M_n}\left(\gamma_n M_n(M_n-1)\right)^{\lambda}\Bigg(\sum_{\x}\sum_{\x'}Q^n(\x)Q^n(\x') \bigg(\sum_{\y}\sqrt{ W^n(\y|\x,s_0)  W^n(\y|\x',s_0)}\bigg)^\frac{1}{\lambda}  \Bigg) ^{\lambda},
\label{eqn:trc_mem_final_example_0}
\end{align}
where the initial state $s_0$ is known at the receiver.
Similarly to what done in \cite[Sec.~5.9]{gallagerBook} to derive the \ac{RCE}, let us introduce the following conditional probability function:
\begin{equation}\label{eqn:trc_mem_final_example_2}
W^n(\y,\s|\x,s_0)=
\begin{cases}
W^n(\y|\x,s_0) \ \text{ if } &\s=\s(\y,s_0)\\
0 \ &\text{ otherwise}.
\end{cases}
\end{equation}
Using \eqref{eqn:trc_mem_final_example_2} in \eqref{eqn:trc_mem_final_example_0} we have:
\begin{align}
P_e(\mathcal{C}_n, s_0) &\leq 
\frac{1}{M_n}\left(\gamma_n M_n(M_n-1)\right)^{\lambda}\Bigg(\sum_{\x}\sum_{\x'}Q^n(\x)Q^n(\x') \bigg(\sum_{\s}\sum_{\y}\sqrt{ W^n(\y,\s|\x,s_0)  W^n(\y,\s|\x',s_0)}\bigg)^\frac{1}{\lambda}  \Bigg) ^{\lambda}\notag\\
&=
\frac{1}{M_n}\left(\gamma_n M_n(M_n-1)\right)^{\lambda}\Bigg(\sum_{\x}\sum_{\x'}Q^n(\x)Q^n(\x') \bigg(\sum_{\s}\prod_{i=1}^n\sum_{y}\sqrt{ W^n(y,s_i|x_i,s_{i-1})  W^n(y,s_i|x_i',s_{i-1})}\bigg)^\frac{1}{\lambda}  \Bigg) ^{\lambda}.
\label{eqn:trc_mem_final_example_4}
\end{align}
Note that in \eqref{eqn:trc_mem_final_example_4} only the state sequence $\s$ corresponding to the initial state $s_0$ and output sequence $\y$ actually leads to a non-zero term in the sum over $\s$. Similarly to \cite[Sec.~5.9]{gallagerBook}, we express the overall bound as a function of a matrix product.
Let:
\begin{align}\label{eqn:trc_mem_final_example_5_0}
\beta(s,s',x,x')=\sum_y\sqrt{ W(y,s|x,s')  W(y,s|x',s')}
\end{align}
where $s,s'\in\{0,1,\ldots,A-1\}$ and $x,x'\in \mathcal{X}$. We define the following set of matrices 
\begin{align}\label{eqn:trc_mem_general_2}
\boldsymbol{A}_{x,x'}=
\begin{pmatrix}
\beta(0,0,x,x') & \beta(0,1,x,x') & \cdots &  \beta(0,A-1,x,x')\\
\beta(1,0,x,x') & \beta(1,1,x,x') & \cdots &  \beta(1,A-1,x,x')\\
\vdots &\vdots & \cdots & \vdots\\
\beta(A-1,0,x,x') & \beta(A-1,1,x,x') & \cdots &  \beta(A-1,A-1,x,x')\\
\end{pmatrix}.
\end{align}
Using \eqref{eqn:trc_mem_final_example_5_0} and \eqref{eqn:trc_mem_general_2} in \eqref{eqn:trc_mem_final_example_4} we have:
\begin{align}
P_e(\mathcal{C}_n, s_0) %&\leq \frac{1}{M_n}\left(\gamma_n M_n(M_n-1)\right)^{\lambda}\left[\sum_{\x}\sum_{\x'}Q^n(\x)Q^n(\x') \left(\sum_{\s}\prod_{i=1}^n\sum_{y}\sqrt{ W(y,s_i|x_i,s_{i-1})  W(y,s_i|x_i,s_{i-1})}\right)^\frac{1}{\lambda}  \right] ^{\lambda}\label{eqn:trc_mem_final_example_6_}\\
&\leq \frac{1}{M_n}\left(\gamma_n M_n(M_n-1)\right)^{\lambda}\Bigg(\sum_{\x}\sum_{\x'}Q^n(\x)Q^n(\x') \bigg(\boldsymbol{e}(s_0)^T \bigg( \prod_{i=1}^n\boldsymbol{A}_{x_i,x_i'} \bigg) \boldsymbol{1}\bigg)^\frac{1}{\lambda}  \Bigg) ^{\lambda}\label{eqn:trc_mem_final_example_6}
%\\
%&=\frac{1}{M_n}\left(\gamma_n M_n(M_n-1)\right)^{\lambda}\left[\sum_{\x}\sum_{\x'}Q^n(\x)Q^n(\x') \left(\boldsymbol{e}(s_0)^T\Pmat^n_{\x,\x'}\boldsymbol{1}\right)^\frac{1}{\lambda}  \right] ^{\lambda}\label{eqn:trc_mem_final_example_7}
\end{align}
where $\boldsymbol{e}(s_0)$ is an $A\times 1$ vector with a $1$ in position $s_0$ and $0$ elsewhere, while $\boldsymbol{1}$ is a $A\times 1$ vector of all $1$'s. 

Finding a single-letter expression for \eqref{eqn:trc_mem_final_example_6} is challenging due to the presence of the matrix product, but can be evaluated using Monte Carlo. We do this in the following numerical example, in which the two-state  model presented in \cite[Fig.~5.9.1]{gallagerBook} is considered.  At a given  instant, the channel is modelled as a \ac{BSC}. The channel transition probability at time $t$ is $p_0$ if the channel was in state $0$ at $t-1$, and $p_1$ otherwise, the probability of state change being $q$. As in \cite[Figure 5.9.2]{gallagerBook}, we assume that the receiver knows the state of the channel at every instant.
\begin{figure*} 
\begin{tabular}{ccc}
\hspace{-1.5em}\tikzset{new spy style/.style={spy scope={%
 magnification=3,
 %size=2cm, 
 connect spies,
 every spy on node/.style={
   rectangle,
   draw,
   },
 every spy in node/.style={
   draw,
   rectangle,
   }
  }
 }
}

\pgfplotsset{scaled x ticks=false}

\begin{tikzpicture}[new spy style]

\footnotesize

\begin{axis}[
title={(a) all rates},
width=3.4in,
height=2.75in,
xmin=0,
xmax=0.5,
ymin=0,
ymax=0.5,
xlabel=$R$,
y label style={at={(axis description cs:0.06,.5)},rotate=0,anchor=south},
ylabel={$E_{\rm trc,lb}(R,Q^n)$}
]

%\addplot[black, line width=2pt] table {
%0	0.26487
%0.0001	0.26019
%0.00025 0.25734
%0.0005  0.25424
%0.001	0.25
%0.002   0.2443
%0.003   0.24018
%0.004   0.23683
%0.005	0.23406
%0.006	0.2317
%0.007	0.22954
%0.008	0.22776
%0.009	0.22584
%0.01	0.22439
%0.011	0.22312
%0.012	0.22151
%0.013	0.2205
%0.014	0.21907
%0.015	0.21831
%0.016	0.21724
%0.017	0.2166
%0.018	0.21482
%0.019	0.21435
%0.02	0.21283
%0.03    0.20313
%0.04    0.19211
%0.05    0.18348
%0.06    0.17296
%0.07    0.16313
%0.08    0.15211
%0.09    0.14348
%0.1     0.13296
%0.11    0.12313
%0.12    0.11211
%0.13    0.10348
%0.14    0.092963
%0.15    0.083134
%0.16    0.072111
%0.17    0.063483
%0.18    0.052963
%0.19    0.042625
%0.2     0.033134
%0.21    0.022111
%0.22    0.013483
%0.23    0.0029633
%0.2328	0
%0.24    0
%0.25	0
%0.26	0
%0.27	0
%0.28	0
%0.29	0
%0.3		0
%0.31	0
%0.32	0
%0.33	0
%0.34	0
%0.35	0
%0.36	0
%0.37	0
%0.38	0
%0.39	0
%0.4		0
%0.41	0
%0.42	0
%0.43	0
%0.44	0
%0.45	0
%0.46	0
%0.47	0
%0.48	0
%0.49	0
%0.5		0
%};
%\addlegendentry{FSC TRC};

\addplot[line width = 2pt, black] table {figures/out_FSC_max.txt};
\addlegendentry{FSC};

%\addplot[line width = 2pt, gray, dashed] table {ISIT2022_simulations/out_memoryless_trc_fine.txt};
%\addlegendentry{DMC TRC};

\addplot[line width = 2pt, gray] table {figures/out_memoryless_max.txt};
\addlegendentry{DMC};

\end{axis}

\end{tikzpicture} & \hspace{-1.5em}
\tikzset{new spy style/.style={spy scope={%
 magnification=3,
 %size=2cm, 
 connect spies,
 every spy on node/.style={
   rectangle,
   draw,
   },
 every spy in node/.style={
   draw,
   rectangle,
   }
  }
 }
} 

\pgfplotsset{scaled x ticks=false}

\begin{tikzpicture}[new spy style]

\footnotesize

\begin{axis}[
title={(b) low rates},
width=1.7in,
height=2.75in,
xmin=0,
xmax=0.02,
ymin=0.2,
ymax=0.45,
xlabel=$R$,
xtick={0,0.01,0.02},
xticklabels={$0$,$0.01$,$0.02$},
y label style={at={(axis description cs:0.15,.5)},rotate=0,anchor=south},
]

%\addplot[black, line width=2pt] table {
%0	0.26487
%0.0001	0.26019
%0.00025 0.25734
%0.0005  0.25424
%0.001	0.25
%0.002   0.2443
%0.003   0.24018
%0.004   0.23683
%0.005	0.23406
%0.006	0.2317
%0.007	0.22954
%0.008	0.22776
%0.009	0.22584
%0.01	0.22439
%0.011	0.22312
%0.012	0.22151
%0.013	0.2205
%0.014	0.21907
%0.015	0.21831
%0.016	0.21724
%0.017	0.2166
%0.018	0.21482
%0.019	0.21435
%0.02	0.21283
%0.03    0.20313
%0.04    0.19211
%0.05    0.18348
%0.06    0.17296
%0.07    0.16313
%0.08    0.15211
%0.09    0.14348
%0.1     0.13296
%0.11    0.12313
%0.12    0.11211
%0.13    0.10348
%0.14    0.092963
%0.15    0.083134
%0.16    0.072111
%0.17    0.063483
%0.18    0.052963
%0.19    0.042625
%0.2     0.033134
%0.21    0.022111
%0.22    0.013483
%0.23    0.0029633
%0.2328	0
%0.24    0
%0.25	0
%0.26	0
%0.27	0
%0.28	0
%0.29	0
%0.3		0
%0.31	0
%0.32	0
%0.33	0
%0.34	0
%0.35	0
%0.36	0
%0.37	0
%0.38	0
%0.39	0
%0.4		0
%0.41	0
%0.42	0
%0.43	0
%0.44	0
%0.45	0
%0.46	0
%0.47	0
%0.48	0
%0.49	0
%0.5		0
%};
%\addlegendentry{FSC TRC};

\addplot[line width = 2pt, black] table {figures/out_FSC_max.txt};
%\addlegendentry{FSC};

%\addplot[line width = 2pt, gray, dashed] table {ISIT2022_simulations/out_memoryless_trc_fine.txt};
%\addlegendentry{DMC TRC};

\addplot[line width = 2pt, gray] table {figures/out_memoryless_max.txt};
%\addlegendentry{DMC};

\end{axis}

\end{tikzpicture}  & \hspace{-2em}\tikzset{new spy style/.style={spy scope={%
 magnification=3,
 %size=2cm, 
 connect spies,
 every spy on node/.style={
   rectangle,
   draw,
   },
 every spy in node/.style={
   draw,
   rectangle,
   }
  }
 }
} 

\pgfplotsset{scaled y ticks=false}

\begin{tikzpicture}[new spy style]

\footnotesize

\begin{axis}[
title=(c) high rates,
width=1.7in,
height=2.75in,
xmin=0.3,
xmax=0.5,
ymin=0,
ymax=0.01,
xlabel=$R$,
ytick={0,0.002,0.004,0.006,0.008,0.01},
yticklabels={$0$,$0.002$,$0.004$,$0.006$,$0.008$,$0.01$},
]

%\addplot[black, line width=2pt] table {
%0	0.26487
%0.0001	0.26019
%0.00025 0.25734
%0.0005  0.25424
%0.001	0.25
%0.002   0.2443
%0.003   0.24018
%0.004   0.23683
%0.005	0.23406
%0.006	0.2317
%0.007	0.22954
%0.008	0.22776
%0.009	0.22584
%0.01	0.22439
%0.011	0.22312
%0.012	0.22151
%0.013	0.2205
%0.014	0.21907
%0.015	0.21831
%0.016	0.21724
%0.017	0.2166
%0.018	0.21482
%0.019	0.21435
%0.02	0.21283
%0.03    0.20313
%0.04    0.19211
%0.05    0.18348
%0.06    0.17296
%0.07    0.16313
%0.08    0.15211
%0.09    0.14348
%0.1     0.13296
%0.11    0.12313
%0.12    0.11211
%0.13    0.10348
%0.14    0.092963
%0.15    0.083134
%0.16    0.072111
%0.17    0.063483
%0.18    0.052963
%0.19    0.042625
%0.2     0.033134
%0.21    0.022111
%0.22    0.013483
%0.23    0.0029633
%0.2328	0
%0.24    0
%0.25	0
%0.26	0
%0.27	0
%0.28	0
%0.29	0
%0.3		0
%0.31	0
%0.32	0
%0.33	0
%0.34	0
%0.35	0
%0.36	0
%0.37	0
%0.38	0
%0.39	0
%0.4		0
%0.41	0
%0.42	0
%0.43	0
%0.44	0
%0.45	0
%0.46	0
%0.47	0
%0.48	0
%0.49	0
%0.5		0
%};
%\addlegendentry{FSC TRC};

\addplot[line width = 2pt, black] table {figures/out_FSC_max.txt};
%\addlegendentry{FSC};

%\addplot[line width = 2pt, gray, dashed] table {ISIT2022_simulations/out_memoryless_trc_fine.txt};
%\addlegendentry{DMC TRC};

\addplot[line width = 2pt, gray] table {figures/out_memoryless_max.txt};
%\addlegendentry{DMC};

\end{axis}

\end{tikzpicture}
\end{tabular}	
 \caption{\ac{TRC} exponent for the \ac{FSC} described in \cite[Sec.~5.9]{gallagerBook} with $q=0.01$, $p_0=0.01$ and $p_1=0.1$ obtained with Monte Carlo for $n=200$ and $10^6$ iterations. The \ac{TRC} over \ac{DMC} is also shown for comparison. The latter is obtained by setting the state transition probability $q$ to $1/2$.}\label{fig:fsc_compare}
 \end{figure*}
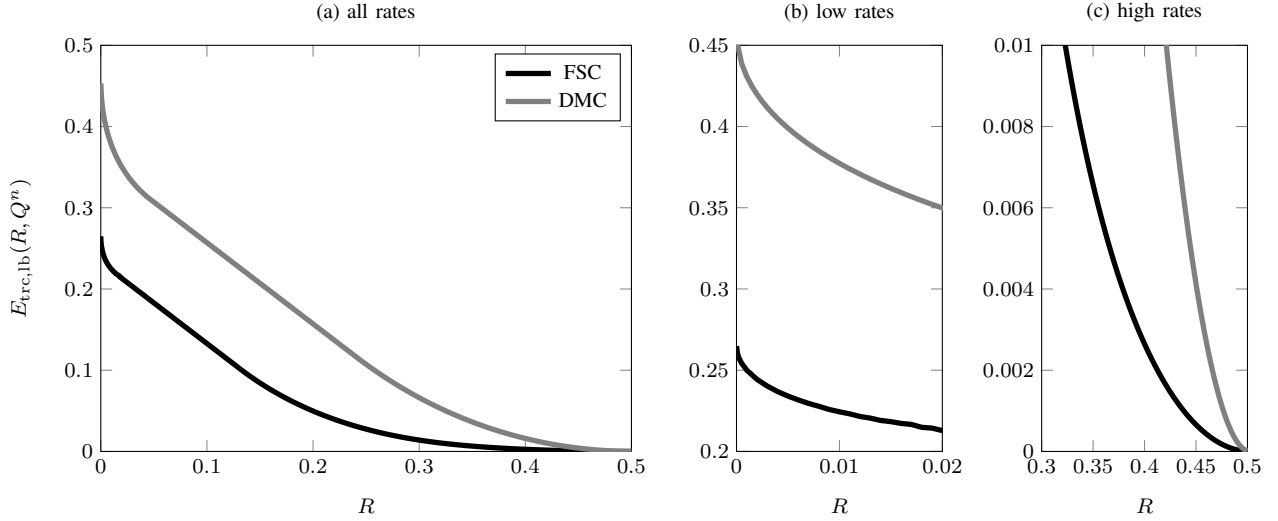
For this specific channel, $\boldsymbol{A}_{x,x'}$ in~\eqref{eqn:trc_mem_general_2} can take two possible values, given by
\begin{align}
\boldsymbol{A}_{0,0}=\boldsymbol{A}_{1,1}=
\begin{pmatrix}
1-q & q\\
q & 1-q
\end{pmatrix},
\end{align}
while:
\begin{align}\label{eqn:trc_mem_final_example_5_2}
\boldsymbol{A}_{0,1}=\boldsymbol{A}_{1,0}=
\begin{pmatrix}
2(1-q)\sqrt{(1-p_0)p_0} & 2q\sqrt{(1-p_1)p_1}\\
2q\sqrt{(1-p_0)p_0} & 2(1-q)\sqrt{(1-p_0)p_0}
\end{pmatrix}.
\end{align}

The bound in \eqref{eqn:theo_lim_finite_sta1} is plotted against the rate in Fig. \ref{fig:fsc_compare}. The term $E_{\rm r}(R,Q)$ given in \eqref{eqn:fsc_gallager} is evaluated in closed form as in \cite[Fig.~5.9.2]{gallagerBook}, while $\Etrclbl$, i.e., the the negative normalized logarithm of \eqref{eqn:trc_mem_final_example_6}, is evaluated using the Monte Carlo method with $10^6$ iterations for a codeword length of $n=200$. More specifically, due to channel memory, the statistical mean inside square brackets in Eq.~\eqref{eqn:trc_mem_final_example_6} cannot be written as a single-letter product, and can only be estimated by randomly generating a large-enough set of pairs of codewords $(\x,\x')$ of length $n$. The \ac{TRC} for the \ac{DMC} channel derived from this FSC (i.e., setting $q=1/2$, see \cite[Sec.~5.9]{gallagerBook}) is also shown for comparison. The result is consistent with the behaviour of the \ac{RCE} presented in \cite{gallagerBook}. 
Note that, for this channel, the memory does not decrease the capacity with respect to its i.i.d.~counterpart. This can be seen in Fig. \ref{fig:fsc_compare}(c) noting that the smallest rates for which the two curves are zero coincide.

\newpage

\appendix

\subsection{Proof of Lemma \ref{lem:591}}
\label{ap:A}
Let us split the length-$n$ input sequence $\x$ into two sequences of length $k$ and $l$ $\x_1$ and $\x_2$, respectively, i.e., $\x=(\x_1,\x_2)$. Let us do the same for the output sequence, i.e., $\y=(\y_1,\y_2)$. Let us also impose that the input distribution be $Q^n(\x)=Q^n(\x_1)Q^n(\x_2)$. This choice can be suboptimal in some cases, but does not impact our proof. With this assumptions and from \eqref{eqn:memo_init_state} it follows that
\begin{align}\label{eqn:591_b}
e^{-n F_{\rm x}^n(\lambda,Q^n)} =& \left(\sum_{\x}\sum_{\x'}Q^n(\x)Q^n(\x') \left(\sum_{\y}\sqrt{\sum_{s_0=1}^A  W^n(\y|\x,s_0)\sum_{s_0'=1}^A  W^n(\y|\x',s_0')}\right)^\frac{1}{\lambda}  \right) ^{\lambda}\\
=&\left(\sum_{\x_1}\sum_{\x_2}\sum_{\x_1'}\sum_{\x_2'}Q^n(\x_1)Q^n(\x_2)Q^n(\x_1')Q^n(\x_2') \left(\sum_{\y_1}\sum_{\y_2}\right.\right.\notag\\\label{eqn:591_c}
&\left.\left.\sqrt{\sum_{s_0=1}^A \sum_{s_k=1}^{A}W^k(\y_1,s_k|\x_1,s_0)W^l(\y_2|\x_2,s_k)\sum_{s_0'=1}^A  \sum_{s_k'=1}^{A}W^k(\y_1,s_k'|\x_1',s_0')W^l(\y_2|\x_2',s_k')}\right)^\frac{1}{\lambda}  \right) ^{\lambda}
\\
=&\left(\sum_{\x_1,\x_1'}Q^n(\x_1)Q^n(\x_1')\sum_{\x_2,\x_2'}Q^n(\x_2)Q^n(\x_2') \left(\sum_{\y_1}\sum_{\y_2}\right.\right.\notag\\\label{eqn:591_d}
&\left.\left.\sqrt{\sum_{s_k=1}^{A} \sum_{s_k'=1}^{A}W^l(\y_2|\x_2,s_k)W^l(\y_2|\x_2',s_k')\sum_{s_0=1}^A\sum_{s_0'=1}^A W^k(\y_1,s_k|\x_1,s_0)W^k(\y_1,s_k'|\x_1',s_0')}\right)^\frac{1}{\lambda}  \right) ^{\lambda}
\\
\leq &\left(\sum_{\x_1,\x_1'}Q^n(\x_1)Q^n(\x_1')\sum_{\x_2,\x_2'}Q^n(\x_2)Q^n(\x_2') \left(\sum_{\y_1}\sum_{\y_2}\right.\right.\notag\\\label{eqn:591_e}
&\left.\left.\sqrt{\sum_{s_k=1}^{A} \sum_{s_k'=1}^{A}W^l(\y_2|\x_2,s_k)W^l(\y_2|\x_2',s_k')\sum_{s_0=1}^A\sum_{s_0'=1}^A W^k(\y_1|\x_1,s_0)W^k(\y_1|\x_1',s_0')}\right)^\frac{1}{\lambda}  \right) ^{\lambda}\\
=&\left(\sum_{\x_1,\x_1'}Q^n(\x_1)Q^n(\x_1') \left(\sum_{\y_1}\sqrt{\sum_{s_0=1}^A\sum_{s_0'=1}^A W^k(\y_1|\x_1,s_0)W^k(\y_1|\x_1',s_0')}\right)^\frac{1}{\lambda}  \right) ^{\lambda}\cdot\notag\\\label{eqn:591_f}
\cdot&
\left(\sum_{\x_2,\x_2'}Q^n(\x_2)Q^n(\x_2') \left(\sum_{\y_2}\sqrt{\sum_{s_k=1}^{A} \sum_{s_k'=1}^{A}W^l(\y_2|\x_2,s_k)W^l(\y_2|\x_2',s_k')}\right)^\frac{1}{\lambda}  \right) ^{\lambda}\\
=&e^{-k F_{k}(\lambda,Q^k)}e^{-l F_{l}(\lambda,Q^l)}
\end{align}
where in \eqref{eqn:591_c} we applied \eqref{eqn:state}, in \eqref{eqn:591_d} we reorganized \eqref{eqn:591_c}, while in \eqref{eqn:591_e} we used the fact that 
\begin{align}
W^k(\y_1,s_k|\x_1,s_0)W^k(\y_1,s_k'|\x_1',s_0') &\leq \sum_{s_k,s_k'}W^k(\y_1,s_k|\x_1,s_0)W^k(\y_1,s_k'|\x_1',s_0')\\
&=W^k(\y_1|\x_1,s_0)W^k(\y_1|\x_1',s_0')
\end{align}
which concludes the proof.

\subsection{Proof of Lemma \ref{lemma:4}}
\label{ap:B}

Let us evaluate $F_{\rm x}^n(\lambda,Q^n)$ at $\lambda=1$. From~\eqref{eqn:ex_s0} we obtain that %\footnote{In \cite{gallagerBook} it is conjectured that $G(\rho,Q^n)$ is the reliability function for $R$ close to capacity [REMOVE?]}:
\begin{align}\label{eqn:ex_s0_E0}
e^{-n F_{\rm x}^n(1,Q^n)} &= \sum_{\x}\sum_{\x'}Q^n(\x)Q^n(\x') \sum_{\y}\sqrt{\sum_{s_0=1}^A  W^n(\y|\x,s_0)\sum_{s_0'=1}^A  W^n(\y|\x',s_0')}  \\
&= \sum_{\y}\sum_{\x}\sum_{\x'}Q^n(\x)Q^n(\x')\sqrt{\sum_{s_0=1}^A  W^n(\y|\x,s_0)\sum_{s_0'=1}^A  W^n(\y|\x',s_0')}\\
&= \sum_{\y}\left(\sum_{\x}Q^n(\x) \sqrt{\sum_{s_0=1}^A  W^n(\y|\x,s_0)}\right)^2\\
&\leq \sum_{\y}\left(\sum_{s_0=1}^A\frac{A}{A}\sum_{\x}Q^n(\x) \sqrt{  W^n(\y|\x,s_0)}\right)^2\\
&= A^2\sum_{\y}\left(\sum_{s_0=1}^A\frac{1}{A}\sum_{\x}Q^n(\x) \sqrt{  W^n(\y|\x,s_0)}\right)^2\\
&\leq A^2\sum_{s_0=1}^A\frac{1}{A}\sum_{\y}\left(\sum_{\x}Q^n(\x) \sqrt{  W^n(\y|\x,s_0)}\right)^2\\
&= A\sum_{s_0=1}^A\sum_{\y}\left(\sum_{\x}Q^n(\x) \sqrt{  W^n(\y|\x,s_0)}\right)^2\\
&\leq A^2\max_{s_0}\sum_{\y}\left(\sum_{\x}Q^n(\x) \sqrt{  W^n(\y|\x,s_0)}\right)^2\\
&=A e^{-n F_0^n(1,Q^n)}.
\end{align}
On the other hand,
\begin{align}\label{eqn:ex_s0_E0}
e^{-n F_{\rm x}^n(1,Q^n)}&= \sum_{\y}\left(\sum_{\x}Q^n(\x) \sqrt{\sum_{s_0=1}^A  W^n(\y|\x,s_0)}\right)^2\\
&\geq \max_{s_0}\sum_{\y}\left(\sum_{\x}Q^n(\x) \sqrt{W^n(\y|\x,s_0)}\right)^2\\
&=\frac{1}{A} e^{-n F_0^n(1,Q^n)}
\end{align}
and, since $A$ is finite, we have that 
\begin{align}\label{eqn:ex_s0_E0_1}
F_0^n(1,Q^n) - \frac{\log A}{n} \leq F_{\rm x}^n(1,Q^n)\leq F_0^n(1,Q^n) + \frac{\log A}{n}.
\end{align}
Using \eqref{eqn:ex_s0_E0_1} and the fact that $A$ is finite, $F_{\rm x}^n(1,Q^n)$ coincides asymptotically with $F_0^n(1,Q^n)$.
Finally, since the chains of inequality above hold for any $n$ and $A$ is finite, if $F_0^n(1,Q^n)$ is finite $F_{\rm x}^n(1,Q^n)$ is also finite.

\bibliographystyle{IEEEtran}
\bibliography{IEEEabrv,Template}

\end{document}